\documentclass[sigconf]{acmart}

\allowdisplaybreaks

\usepackage{algorithm,algorithmic}
\usepackage{subcaption}
\usepackage{multirow}
\usepackage{tabularx}
\usepackage{arydshln}
\usepackage{natbib}

\newcolumntype{L}[1]{>{\raggedright\let\newline\\\arraybackslash\hspace{0pt}}m{#1}}

\usepackage{balance}

\def\BibTeX{{\rm B\kern-.05em{\sc i\kern-.025em b}\kern-.08emT\kern-.1667em\lower.7ex\hbox{E}\kern-.125emX}}
    
\copyrightyear{2020}
\acmYear{2020}
\setcopyright{acmcopyright}
\acmConference[WSDM '20]{The Thirteenth ACM International Conference on Web Search and Data Mining}{February 3--7, 2020}{Houston, TX, USA}
\acmBooktitle{The Thirteenth ACM International Conference on Web Search and Data Mining (WSDM '20), February 3--7, 2020, Houston, TX, USA}
\acmPrice{15.00}
\acmDOI{10.1145/3336191.3371854}
\acmISBN{978-1-4503-6822-3/20/02}

\begin{document}

\fancyhead{}

\title[Knowledge-aware Complementary Product]{Knowledge-aware Complementary Product Representation Learning}

\author{Da Xu}
\authornote{Both authors contributed equally to this research.}
\author{Chuanwei Ruan}
\authornotemark[1]
\affiliation{%
  \institution{Walmart Labs}
  \city{Sunnyvale}
  \state{California}
  \country{USA}
}
\email{{Da.Xu,Chuanwei.Ruan}@walmartlabs.com}

\author{Jason Cho, Evren Korpeoglu, Sushant Kumar, Kannan Achan}
\affiliation{%
  \institution{Walmart Labs}
  \city{Sunnyvale}
  \state{California}
  \country{USA}
}
\email{{HCho, EKorpeoglu,SKumar4,KAchan}@walmartlabs.com}
%
\renewcommand{\shortauthors}{Trovato and Tobin, et al.}

%
\begin{abstract}
Learning product representations that reflect complementary relationship plays a central role in e-commerce recommender system. In the absence of the product relationships graph, which existing methods rely on, there is a need to detect the complementary relationships directly from noisy and sparse customer purchase activities. Furthermore, unlike simple relationships such as similarity, complementariness is asymmetric and non-transitive. Standard usage of representation learning emphasizes on only one set of embedding, which is problematic for modelling such properties of complementariness.

We propose using knowledge-aware learning with dual product embedding to solve the above challenges. We encode contextual knowledge into product representation by multi-task learning, to alleviate the sparsity issue. By explicitly modelling with user bias terms, we separate the noise of customer-specific preferences from the complementariness. Furthermore, we adopt the dual embedding framework to capture the intrinsic properties of complementariness and provide geometric interpretation motivated by the classic separating hyperplane theory. Finally, we propose a Bayesian network structure that unifies all the components, which also concludes several popular models as special cases.

The proposed method compares favourably to state-of-art methods, in downstream classification and recommendation tasks. We also develop an implementation that scales efficiently to a dataset with millions of items and customers.

\end{abstract}

%
%

%

%
\maketitle

\section{Introduction}
\label{sec:Introduction}

Modern recommender systems aim at providing customers with personalized contents efficiently through the massive volume of information. Many of them are based on customers' explicit and implicit preferences, where content-based \cite{soboroff1999combining} and collaborative filtering systems \cite{koren2015advances} have been widely applied to e-commerce \cite{linden2003amazon}, online media \cite{saveski2014item} and social network \cite{konstas2009social}. In recent years representation learning methods have quickly gained popularity in online recommendation literature. Alibaba \cite{wang2018billion} and Pinterest \cite{ying2018graph} have deployed large-scale recommender system based on their trained product embeddings. Youtube also use their trained video embeddings as part of the input to a deep neural network \cite{covington2016deep}. Different from many other use cases, e-commerce platforms often offer a vast variety of products, and nowadays customers shop online for all-around demands from electronics to daily grocery, rather than specific preferences on narrow categories of products. Therefore understanding the intrinsic relationships among products \cite{zheng2009substitutes} while taking care of individual customer preferences motivates our work. A topic modelling approach was recently proposed to infer complements and substitutes of products as link prediction task using the extracted product relation graphs of Amazon.com \cite{mcauley2015inferring}.

For e-commerce, product complementary relationship is characterized by co-purchase patterns according to customer activity data. For example, many customers who purchase a new \textsl{TV} often purchase \textsl{HDMI cables} next and then purchase \textsl{cable adaptors}. Here \textsl{HDMI cables} are complementary to \textsl{TV}, and \textsl{cable adaptors} are complementary to \textsl{HDMI cable}. We use $\rightarrow$ as a shorthand notation for 'complement to' relationship. By this example, we motivate several properties of the complementary relationship:
\begin{itemize}
\item 
\textbf{Asymmetric}. \textsl{HDMI cable} $\rightarrow$ \textsl{TV}, but \textsl{TV} $\not \rightarrow$ HDMI cable.
\item 
\textbf{Non-transitive}. Though \textsl{HDMI cables} $\rightarrow$ \textsl{TV}, and \textsl{cable adaptors} $\rightarrow$ \textsl{HDMI cable}, but \textsl{cable adaptors} $\not \rightarrow$ \textsl{TV}.
\item 
\textbf{Transductive}. \textsl{HDMI cables} are also likely to complement other \textsl{TVs} with similar model and brand. 
\item
\textbf{Higher-order}. Complementary products for meaningful product combos, such as (\textsl{TV}, \textsl{HDMI cable}), are often different from their individual complements.
\end{itemize}

Although the above properties for complementary relationship highlight several components for designing machine learning algorithms, other manipulations are also needed to deal with the \textbf{noise} and \textbf{sparsity} in customer purchase activity data. Firstly, the low signal-to-noise ratio in customer purchase sequences causes difficulty in directly extracting complementary product signals from them. Most often, we only observe a few complementary patterns in customer purchase sequences or baskets. 
We list a customer's single day purchase as a sequence for illustration:

\{\textsl{Xbox}, \textsl{games}, \textsl{T-shirt}, \textsl{toothbrush}, \textsl{pencil}, \textsl{notepad}\}.


Notice that among the fifteen pairs of possible item combinations, only two paris can be recognized as complementary, i.e \textsl{games} $\rightarrow$ \textsl{Xbox}, \textsl{notepad} $\rightarrow$ \textsl{pencil}. The rest purchases are out of the customer's personal interest.
As a solution to the noise issue, we introduce a customer-product interaction term to directly take account of the noises, while simultaneously modelling personalized preferences. We use $\mathcal{U}$ to denote the set of users (customers) and $\mathcal{I}$ to denote the set of items (products). Let $\mathbf{U}\in \mathcal{U}$ be a user categorical random variable and $\{\mathbf{I}_{t-1},\dots,\mathbf{I}_{t-k}\}$ be a sequence of item categorical random variables that represents $k$ consecutive purchases before time $t$. If we estimate the conditional probability $p(\mathbf{I}_{t+1}|\mathbf{U},\mathbf{I}_t,\dots,\mathbf{I}_{t-k})$ with \textsl{softmax} classifier under score function $S(.)$, then previous arguments suggest that $S\big(\mathbf{I}_{t+1},(\mathbf{U},\mathbf{I}_t,\dots,\mathbf{I}_{t-k})\big)$ should consist of a user-item preference term and a item complementary pattern term:
\begin{equation}
\label{eqn:decomposition}
     S\big(\mathbf{I}_{t+1},(\mathbf{U},\mathbf{I}_t,\dots,\mathbf{I}_{t-k})\big) = f_{UI}(\mathbf{I}_{t+1},\mathbf{U})+f_{I}(\mathbf{I}_{t+1},\mathbf{I}_t,\dots,\mathbf{I}_{t-k}).
\end{equation}
Now we have $f_{UI}(.)$ accounting for user-item preference (bias) and $f_I(.)$ characterizing the strength of item complementary pattern. When the complementary pattern of an purchase sequence is weak, i.e $f_{I}(\mathbf{I}_{t+1},\mathbf{I}_t,\cdots,\mathbf{I}_{t-k})$ is small, the model will enlarge the user-item preference term, and vice versa.

On the other hand, the sparsity issue is more or less standard for recommender systems and various techniques have been developed for both content-based and collaborative filtering systems \cite{papagelis2005alleviating,huang2004applying}. Specifically, it has been shown that modelling with contextual information boosts performances in many cases \cite{melville2002content, balabanovic1997fab, hannon2010recommending}, which also motivates us to develop our context-aware solution.

Representation learning with shallow embedding gives rises to several influential works in natural language processing (NLP) and geometric deep learning. The \textsl{skip-gram} (\textsl{SG}) and \textsl{continuous bag of words} (\textsl{CBOW}) models \cite{mikolov2013distributed} as well as their variants including \textsl{GLOVE} \cite{pennington2014glove}, \textsl{fastText} \cite{bojanowski2017enriching}, \textsl{Doc2vec} (\textsl{paragraph vector}) \cite{le2014distributed} have been widely applied to learn word-level and sentence-level representations. While classical node embedding methods such as \textsl{Laplacian eigenmaps} \cite{belkin2003laplacian} and \textsl{HOPE} \cite{ou2016asymmetric} arise from deterministic matrix factorization, recent work like \textsl{node2vec} \cite{grover2016node2vec} explore from the stochastic perspective using random walks. However, we point out that the word and sentence embedding models target at semantic and syntactic similarities while the node embedding models aim at topological closeness. These relationships are all symmetric and mostly transitive, as opposed to the complementary relationship. Furthermore, the \textbf{transductive} property of complementariness requires that similar products should have similar complementary products, suggesting that product similarity should also be modelled explicitly or implicitly. We approach this problem by encoding contextual knowledge to product representation such that contextual similarity is preserved.

We propose the novel idea of using context-aware dual embeddings for learning complementary products. While both sets of embedding are used to model complementariness, product similarities are implicitly represented on one of the embeddings by encoding contextual knowledge. Case studies and empirical testing results on the public \textsl{Instacart} and a proprietary e-commerce dataset show that our dual product embeddings are capable of capturing the desired properties of complementariness, and achieve cutting edge performance in classification and recommendation tasks.

\section{Contributions and Related Works}
\label{sec:related-works}

Compared to the previously published work on learning product representations and modelling product relationships for e-commerce, our contributions are summarized below.

\textbf{Learning higher-order product complementary relationship with customer preferences} - Since product complementary patterns are mostly entangled with customer preference, we consider both factors in our work. And by directly modelling whole purchase sequences $f_{I}(\mathbf{I}_{t+1},\mathbf{I}_t,\cdots,\mathbf{I}_{t-k})$, we are able to capture \textbf{higher-order} complementary relationship. The previous work of inferring complements and substitutes with topic modelling \cite{mcauley2015inferring} and deep neural networks \cite{Zhang:2018:QNC:3240323.3240368} on Amazon data relies on the extracted graph of product relationships. Alibaba proposes an approach by first constructing the weighted product co-purchase count graph from purchase records and then implementing a node embedding algorithm \cite{wang2018billion}. However, individual customer information is lost after the aggregation. The same issue occurs in item-based collaborative filtering  \cite{sarwar2001item} and product embedding \cite{vasile2016meta}. A recent work on learning grocery complementary relationship for next basket predict models (item, item, user) triplets extracted from purchases sequences, and thus do not account for higher-order complementariness \cite{wan2018representing}.

\textbf{Use context-aware dual product embedding to model complementariness} - Single embedding space may not capture \textbf{asymmetric} property of complementariness, especially when they are treated as projections to lower-dimensional vector spaces where inner products are symmetric. Although Youtube's approach takes both user preference and video co-view patterns into consideration, they do not explore complementary relationship \cite{covington2016deep}. PinSage, the graph convolutional neural network recommender system of Pinterest, focus on symmetric relations between their pins \cite{ying2018graph}. The triplet model \cite{wan2018representing} uses two sets of embedding, but contextual knowledge is not considered. To our best knowledge, we are the first to use context-aware dual embedding in modelling product representations.

\textbf{Propose a Bayesian network structure to unify all components, and conclude several classic models as special cases} - As we show in Table \ref{tab:pgm}, the classic collaborative filtering models for product recommendation \cite{linden2003amazon}, sequential item models such as \textsl{item2vec} \cite{barkan2016item2vec} and \textsl{metapath2vec} \cite{dong2017metapath2vec} for learning product representation, the recent \textsl{triplet2vec} model \cite{wan2018representing} as well as the \textsl{prod2vec} \cite{grbovic2015commerce} model which considers both purchase sequence and user, can all be viewed as special cases within the Bayesian network representation of our model.

Besides the above major contributions, we also propose a fast inference algorithm to deal with the \textbf{cold start} challenge \cite{lam2008addressing,schein2002methods}, which is crucial for modern e-commerce platforms. Different from the cold-start solutions for collaborative filtering using matrix factorization methods \cite{zhou2011functional,bobadilla2012collaborative}, we infer from product contextual features. The strategy is consistent with recent work, which finds contextual features playing an essential role in mitigating the cold-start problem \cite{saveski2014item,gantner2010learning}. We also show that cold-start product representations inferred from our algorithm empirically perform better than simple context similarity methods in recommendation tasks.

\section{Method}
\label{sec:method}

In this section, we introduce the technical details of our method. We first describe the Bayesian network representation of our approach, specify the factorized components and clarify their relationships. We then introduce the loss function in Section \ref{sec:prob_model}. In Section \ref{sec:parama}, we define the various types of embeddings and how we use them to parameterize the score functions. We then discuss two ranking criteria for recommendation tasks using embeddings obtained from the proposed approach, in Section \ref{sec:ranking}. The geometric interpretation of using dual product embeddings and user bias term is discussed in Section \ref{sec:interpretation}. Finally, we present our fast inference algorithm for cold-start products in Section \ref{sec:cold-start}.

\subsection{Probability model for purchase sequences}
\label{sec:prob_model}

\begin{table}[bth]
    \centering
    \begin{tabular}{|p{1.5cm}|c|c|}
    \hline 
     & Special cases & Proposed model \\ \hline 
     Item-item \textsl{CF} models \cite{linden2003amazon} &   \raisebox{-\totalheight}{\includegraphics[scale=0.3]{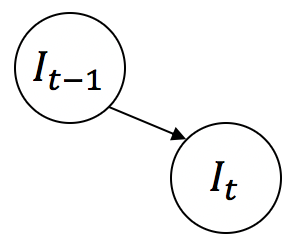}} & \multirow{4}{*}{\raisebox{-1.5\totalheight}{\includegraphics[scale=0.3]{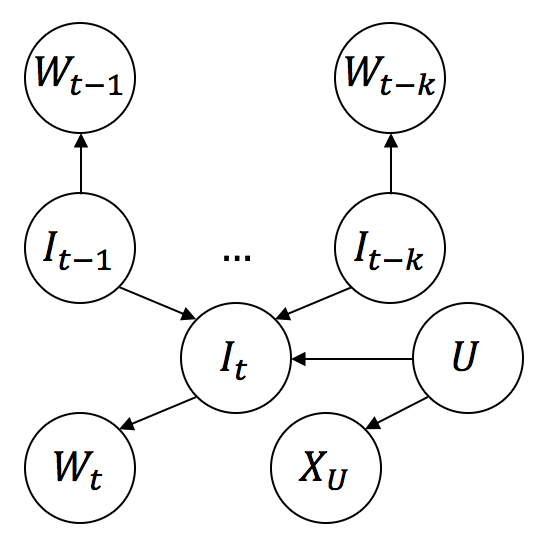}}} \\ \cline{1-2}
     Sequential models (\textsl{item2vec} \cite{barkan2016item2vec}) &  \raisebox{-\totalheight}{\includegraphics[scale=0.3]{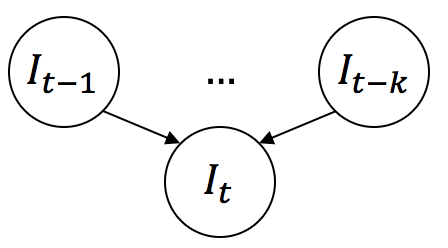}} &  \\ \cline{1-2}
     \textsl{triplet2vec} \cite{wan2018representing} & \raisebox{-\totalheight}{\includegraphics[scale=0.3]{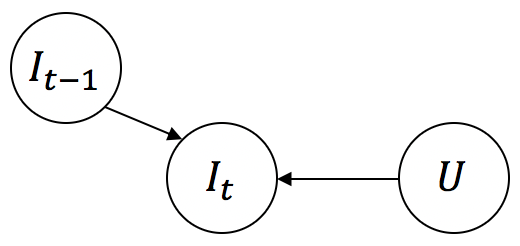}} & \\ \cline{1-2}
     \textsl{prod2vec} \cite{grbovic2015commerce} & \raisebox{-\totalheight}{\includegraphics[scale=0.3]{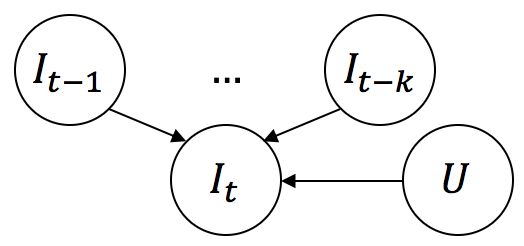}} & \\ \hline
    \end{tabular}
    \caption{\small The Bayesian network representation of joint probability distribution of observation, for our proposed approach (right panel) and classic recommendation algorithms (left panels). }
    \label{tab:pgm}

\end{table}

Let $\mathcal{U}$ be the set of $N$ users and $\mathcal{I}$ be the set of $M$ items. The contextual knowledge features for items such as brand, title and description are treated as tokenized words. Product categories and discretized continuous features are also treated as tokens. Without loss of generality, for each item we concatenate all the tokens into a vector denoted by $\mathbf{w}_i$ for item $i \in \mathcal{I}$. We use $\mathcal{W}$ to denote the whole set of tokens which has $n_w$ instances in total. Similarly, each user $u$ has a knowledge feature vector of tokens $\mathbf{x}_u$ and $\mathcal{X}$ denotes the whole set of $n_x$ user features. A complete observation for user $u$ with the first purchase after time $t$ and previous $k$ ($k$ may vary) consecutive purchases is given by $\{ u, i_{t}, i_{t-1}, \dots, i_{t-k}, \mathbf{x}_u, \mathbf{w}_{i_t}, \mathbf{w}_{i_{t-1}}, \dots, \mathbf{w}_{i_{t-k}} \}$.

\textbf{Bayesian factorization}. The model is optimized for predicting the next purchase according to the user and the most recent purchases. To encode contextual knowledge into item/user representations, we consider the contextual knowledge prediction task:
\begin{itemize}
    \item $p(\mathbf{I}_t | \mathbf{U}, \mathbf{I}_{t-1}, \dots, \mathbf{I}_{t-k})$ - Predict next-to-buy item given user and recent $k$ purchases. 
    
    \item $p(\mathbf{W}_I | \mathbf{I})$ - Predict items' contextual knowledge features. By assuming the conditional independence of contextual features, we adopt the factorization such that:
    $$p(\mathbf{W}_I | \mathbf{I}) = \prod_{\mathbf{W} \in \mathbf{W}_I} p(\mathbf{W} | \mathbf{I}).$$
    
    \item $p(\mathbf{X}_U | \mathbf{U})$ - Predict users' features. We also factorize this term into $\prod_{\mathbf{X} \in \mathbf{X}_U} p(\mathbf{X} | \mathbf{U})$.
\end{itemize}

To put together all the components, we first point out that the joint probability function of the full observation has an equivalent Bayesian network representation \cite{wainwright2008graphical} (right panel of Table \ref{tab:pgm}). The log probability function now has the factorization (decomposition) shown in (\ref{eqn:joint-prob}). Since we do not model the marginals of $p(\mathbf{I})$ and $p(\mathbf{U})$ with embeddings, we treat them as constant terms denoted by $C$ in (\ref{eqn:joint-prob}). 

\begin{equation}
\label{eqn:joint-prob}
    \begin{split}
        &\log p(\mathbf{U}, \mathbf{I}_t, \mathbf{I}_{t-1}, \cdots, \mathbf{I}_{t-k}, \mathbf{X}_U, \mathbf{W}_{I_t}, \mathbf{W}_{I_{t-1}}, \cdots, \mathbf{W}_{I_{t-k}}) \\
        =& \log \big\{ p(\mathbf{I}_t | \mathbf{U}, \mathbf{I}_{t-1}, \cdots, \mathbf{I}_{t-k}) p(\mathbf{X}_U | \mathbf{U}) \prod_{j=0}^{k} \log p(\mathbf{W}_{I_{t-j}} | \mathbf{I}_{t-j})\big\} +C \\
        =& \log p(\mathbf{I}_t | \mathbf{U}, \mathbf{I}_{t-1}, \cdots, \mathbf{I}_{t-k}) \\
        &+ \underbrace{\sum_{\mathbf{X} \in \mathbf{X}_U}\log p(\mathbf{X} | \mathbf{U}) + \sum_{j=0}^k \sum_{\mathbf{W} \in \mathbf{W}_{I_{t-j}}} \log p(\mathbf{W}| \mathbf{I}_{t-j})}_{\text{Contextual knowledge}} + C.
    \end{split}
\end{equation}

\textbf{Log-odds and loss function}. As we have pointed out before, several related models can be viewed as special cases of our approach under the Bayesian network structure. Visual presentations of these models are shown in Table \ref{tab:pgm}. Classic item-item collaborative filtering for recommendation works on conditional probability of item pairs $p(I_t | I_{t-1})$. Sequential item models such as \textsl{item2vec} work on $p(I_{t}| I_{t-1}, \cdots, I_{t-k})$. The recently proposed triplet model focus on (item, item, user) triplet distribution: $p(I_t | I_{t-1}, U)$, and \textsl{prod2vec} models user purchase sequence as $p(I_t | I_{t-1},\cdots,I_{t-k}, U)$.

In analogy to \textsl{SG} and \textsl{CBOW} model, each term in (\ref{eqn:joint-prob}) can be treated as as multi-class classification problem with \textsl{softmax} classifier. However, careful scrutiny reveals that such perspective is not suitable under e-commerce setting. Specifically, using softmax classification function for $p(\mathbf{I_t} | \mathbf{U}, \mathbf{I_{t-1}}, \dots, \mathbf{I_{t-k}})$ implies that given the user and recent purchases we only predict one next-to-buy complementary item. However, it is very likely that there are several complementary items for the purchase sequence. The same argument holds when modelling item words and user features with multi-class classification. Instead, we treat each term in (\ref{eqn:joint-prob}) as the log of probability for the Bernoulli distribution. Now the semantic for $p(\mathbf{I}_t | \mathbf{U}, \mathbf{I}_{t-1}, \dots, \mathbf{I}_{t-k})$ becomes that given the user and recent purchases, whether $I_t$ is purchased next. Similarly, $p(\mathbf{W}| \mathbf{I})$ now implies whether or not the item has the target context words. 

Let $s(i,u)$ models personalized preference of user $u$ on item $i$ and $s(i_t,t_{t-1:t-k})$ measures the high-order item complementariness given k previously purchased items. A larger value of the two score functions implies stronger user preferences and complementary patterns, respectively. Following the decomposition (\ref{eqn:decomposition}), the log-odds of next-to-buy complementary item is given by (\ref{eqn:log_odss}),  

\begin{equation}
\label{eqn:log_odss}
\log\Big(\frac{P(\mathbf{I}_t = i_t | \mathbf{U}, \mathbf{U}_{t-1} \dots \mathbf{I}_{t-k})}{P(\mathbf{I}_t \neq i_t | \mathbf{U}, \mathbf{U}_{t-1} \dots \mathbf{I}_{t-k})}\Big) = s\big(i_t, u\big) + s\big(i_t,t_{t-1:t-k}\big)  .
\end{equation}

Similarly, we define $s(w, i)$ as the relatedness of item feature $w$ and item $i$, $s(x, u)$ as the relatedness of user feature $x$ and user $u$. Therefore, by treating the observed instance as the positive label and all other instances as negative labels, the loss function for each complete observation under binary logistic loss can be formulated according to the log-likelihood function decomposition in (\ref{eqn:joint-prob}):

\begin{equation}
\small
\label{eqn:loss}
    \begin{split}
        & \ell = \log \big(1 + e^{-s(i_t,u) - s(i_t,i_{t-1:t-k})} \big) + \sum_{\tilde{i} \in \mathcal{I} / i_t} \log \big( 1 + e^{s(\tilde{i},u) + s(\tilde{i},i_{t-1:t-k})} \big) \\
        &+ \sum_{j=0}^k \sum_{w \in \mathbf{w}_{i_{t-j}}} \Big\{ \log \big(1 + e^{-s(w,i_{t-j})} \big) + \sum_{\substack{\tilde{w} \\ \in \mathcal{W} \setminus \mathbf{w}_{i_{t-j}}}} \log \big( 1 + e^{s(\tilde{w},i_{t-j})} \big) \Big\} \\
        &+ \sum_{x \in \mathbf{x}_u} \Big\{ \log \big(1 + e^{-s(x,u)} \big) + \sum_{\tilde{x} \in \mathcal{X} \setminus x} \log \big( 1 + e^{s(\tilde{x},u)} \big) \Big\}. \\
    \end{split}
\end{equation}

We notice that binary classification loss requires summing over all possible negative instances, which is computationally impractical. We use negative sampling as an approximation for all binary logistic loss terms, with frequency-based negative sampling schema proposed in \cite{mikolov2013distributed}. For instance, the first line in (\ref{eqn:loss}) is now approximated by (\ref{eqn:negative-sample}) where $\text{Neg}(\mathcal{I})$ denotes a set of negative item samples.

\begin{equation}
\label{eqn:negative-sample}
\log \big(1 + e^{-s(i_t,u) - s(i_t,i_{t-1:t-k})} \big) + \sum_{\tilde{i} \in \text{Neg}(\mathcal{I})} \log \big( 1 + e^{s(\tilde{i},u) + s(\tilde{i},i_{t-1:t-k})} \big)
\end{equation}

\subsection{Parameterization with context-aware dual embeddings}
\label{sec:parama}

In last section we give the definitions of the log-odds function: $s(i, u)$, $s(i_t,(i_{t-1}, \dots, i_{t-k}))$, $s(w, i)$ and $s(x ,u)$. However, exact estimation of all these functions is impractical as the their amount grows quadratically with the number of items and users. Motivated by previous work in distributed representation learning, we embed the item $i$, user $u$, item feature $w$ and user feature $x$ into low dimensional vector space and apply the dot product to approximate the log-odds functions.

\textbf{Item embeddings in dual space}. Let $Z^I \in \mathbb{R}^{M \times P}$ and $\tilde{Z}^I \in \mathbb{R}^{M \times P}$ be the dual item embeddings, such that $s(i_1,i_2) = \langle \tilde{\mathbf{z}}^I_{i_1}, \mathbf{z}^I_{i_2} \rangle$ and $s(i_2,i_1) = \langle \tilde{\mathbf{z}}^I_{i_2}, \mathbf{z}^I_{i_1} \rangle$. We refer to $Z^I$ as item-in embedding and $\tilde{Z}^I$ as item-out embedding. By employing the dual embeddings, the model is able to capture the \textbf{asymmetric} and \textbf{non-transitive} property. Notice that inner products with only one set of embeddings is inherently symmetric and transitive according to their definition in the Euclidean space.

\textbf{Knowledge-aware item embedding and user embedding}. Let $Z^W \in \mathbb{R}^{n_w \times P}$ be the set of word embeddings, such that $s(w,i) = \langle \mathbf{z}^W_w,\mathbf{z}^I_i \rangle$. Similarly, let $Z^X \in \mathbb{R}^{n_x \times P}$ be the set of discretize user feature embeddings such that $s(x,u) = \langle \mathbf{z}^X_x,\mathbf{z}^U_u \rangle$. By imposing the contextual constraints in the loss function (\ref{eqn:loss}), contextually similar items and users are enforced to have embeddings close to each other. Due to the \textbf{transductive} property of the complementariness relationship, imposing the contextual knowledge constraints helps alleviate the sparisty issue since unpopular items could leverage the information from their similar but more popular counterparts. 

\textbf{User-item interaction term}. Let $Z^U \in \mathbb{R}^{N \times P}$ be the set of embeddings for users and $Z^{IU} \in \mathbb{R}^{M \times P}$ be the set of embeddings for item-user context, such that $s(i,u) = \langle \tilde{ \mathbf{z}}^{I}_i,\mathbf{z}^U_u \rangle$. By capturing the user specific bias in the observed data, learning the user-item interactions helps discover a more general and intrinsic complementary pattern between items.

\textbf{Higher-order complementariness}. We then define the score function when the input is an item sequence, i.e $s(i_t,(i_{t-1}, \dots, i_{t-k}))$. Making prediction based on item sequence has also been spotted in other embedding-based recommender systems. PinSage has experimented on mean pooling and max pooling as well as using LSTM \cite{ying2018graph}. Youtube uses mean pooling \cite{covington2016deep}, and another work from Alibaba for learning user interests proposes using the attention mechanism \cite{zhou2018deep}. We choose using simple pooling methods over others for interpretability and speed. Our preliminary experiments show that mean pooling constantly outperforms max pooling, so we settle down to the mean pooling shown in (\ref{eqn:mean-pool}).
\begin{equation}
\label{eqn:mean-pool}
    s(i_t,i_{t-1:t-k}) \equiv s\big(i_t,(i_{t-1}, \dots, i_{t-k})\big) = \frac{ \langle \tilde{\mathbf{z}}^I_{i_t},\sum_{j=1}^k \mathbf{z}^{I}_{i_{t-j}} \rangle}{k}.
\end{equation}
It is now straightforward to see that $s(i,u)$ and $s\big(i_t,(i_{t-1}, \dots, i_{t-k})\big)$ meet the demand of a realization of $f_{UI}(.)$ and $f_I(.)$ in (\ref{eqn:decomposition}).

\subsection{Ranking criteria for recommendation}
\label{sec:ranking}
In this section, we briefly discuss how to conduct online recommendation with the optimized embeddings, where top-ranked items are provided for customers based on their recent purchases and/or personal preferences.

The first criteria only relies on the score function $s(i_t,i_{t-1:t-k})$ given by (\ref{eqn:mean-pool}), which indicates the strength of complementary signal between the previous purchase sequence and target item. 

The second criterion accounts for both user preference and complementary signal strength, where target items are ranked according to $s(i_t,u) + s(i_t,i_{t-1:t-k})$. According to the decomposition in (\ref{eqn:decomposition}), the second ranking criterion can be given by:
\begin{equation}
\label{eqn:rank_with_user}
    S\big(\mathbf{I}_{t+1},(\mathbf{U},\mathbf{I}_t,\dots,\mathbf{I}_{t-k})\big) = \langle \tilde{\mathbf{z}}^I_{i_t}, \mathbf{z}^U_u  \rangle + \frac{ \langle \tilde{\mathbf{z}}^I_{i_t},\sum_{j=1}^k \mathbf{z}^{I}_{i_{t-j}} \rangle}{k}.
\end{equation}

The advantage for the second criterion is that user preferences are explicitly considered. However, it might drive the recommendations away from the most relevant complementary item to compensate for user preference. Also, the computation cost is doubled compared with the first criterion. To combine the strength of both methods, we use the first criterion to recall a small pool of candidate items and then use the second criterion to re-rank the candidates.

\subsection{Geometric Interpretation }
\label{sec:interpretation}

Here we provide our geometric interpretation for the additional item-out embedding $\tilde{Z}^I$ and the user bias term. The use of dual item embedding is not often seen in embedding-based recommender system literature, but it is essential for modelling the intrinsic properties of complementariness. Suppose all embeddings are fixed except $\tilde{Z}^I$. According to classical separating hyperplane theory, the vector $\tilde{\mathbf{z}}_j$ is actually the normal of the separating hyperplane for item $j$ with respect to the embeddings of the positive and negative purchase sequences. In other words, for item $j$ as a next-to-buy complementary item, the hyperplane tries to identify which previous purchase sequences are positive and which are not. 

\begin{figure}[htb]
     \centering

    \begin{minipage}{0.45\textwidth}
        \centering
        \includegraphics[width=0.85\textwidth]{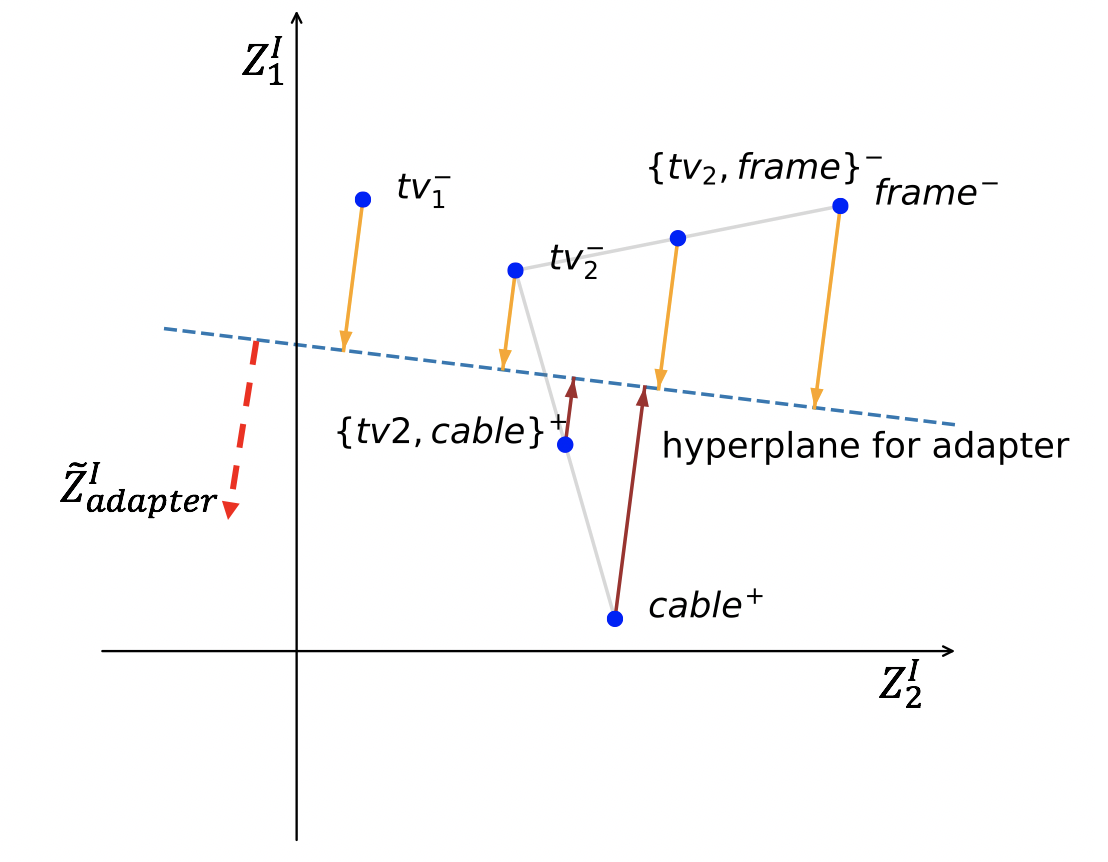} 
        \caption{
            \small Geometric illustration for the separating hyperplane interpretation with $k=1,2$ and $P_I=2$. For \texttt{adaptor} as complementary item, \texttt{TV}$^-$, \texttt{frame}$^-$ are negative samples and \texttt{cable}$^+$ is positive sample. Also, \texttt{(TV$_1$,cable)}$^+$ combo is positive sample and \texttt{(TV$_2$,frame)}$^-$ is negative sample, where combo is represented by the centroid of the two items. $\tilde{z}^I_{\text{adaptor}}$ is optimized such that positive/negative samples have positive/negative distances to the separating hyperplane characterized by its normal vector $\tilde{z}^I_{\text{adaptor}}$ and intercept $b_{\text{adaptor}}$ (not specified here).
        } 
        \label{fig:geometric2}
    \end{minipage}
\end{figure}

Consider the total loss function $L = \sum_{q=1}^n \ell_q$, where $n$ is the number of observed purchase sequences and the subscript $q$ gives the index of the observation. In the loss function (\ref{eqn:loss}), item-out embeddings $\tilde{Z}^I$ only appears in the first two terms. Since $L$ is separable, we collect all terms that involve the item-out embedding of item $j$ as shown in (\ref{eqn:logistic-reg}). For clarity purpose we let $k=1$ in (\ref{eqn:logistic-reg}), i.e only the most recent purchase is included.

\begin{equation}
\small
\label{eqn:logistic-reg}
    \begin{split}
        \mathcal{L}_j = \sum_{(i,u) \in \mathcal{P}} \log \big( 1+e^{-b_{j,u} - \langle \tilde{\mathbf{z}}^I_j,\mathbf{z}^I_i \rangle} \big) + \sum_{(\tilde{i},u) \in \mathcal{N}} \log \big( 1+e^{b_{j,u} + \langle \tilde{\mathbf{z}}^I_j,\mathbf{z}^I_{\tilde{i}} \rangle} \big).
    \end{split}
\end{equation}
In (\ref{eqn:logistic-reg}), $\mathcal{P}$ represents the whole set of item-user $(i,u)$ pairs in the observed two-item purchase sequences $\{j, i\}$ for user $u$. $\mathcal{N}$ denotes other $(\tilde{i},u)$ pairs where item $j$ is used as one of the negative samples in (\ref{eqn:negative-sample}). The scalar $b_{j,u} \equiv s(j,u)$ is the preference of user $u$ on item $j$. It is then obvious that optimizing $ \tilde{\mathbf{z}}^I_j$ in (\ref{eqn:logistic-reg}) is equivalent to solving a logistic regression with $\tilde{\mathbf{z}}^I_j$ as parameters. The design matrix is constructed from the fixed item-in embeddings $\{\mathbf{z}^I_i\}_{(i,u) \in \mathcal{P}\cup \mathcal{N}}$. One difference from ordinary logistic regression is that we have a fixed intercept terms $b_{j,u}$ for each user. Analytically this means that we use the user's preferences on item $j$, i.e. $s(j,u)$, as the intercept, when using his/her purchase sequences to model the complementary relationship between item $j$ and other items. In logistic regressions, the regression parameter vector $\tilde{\mathbf{z}}^I_j$ gives the normal of the optimized separating hyperplane. We provide a sketch of the concept in Figure \ref{fig:geometric2}.

When $k \geq 2$, we replace $\mathbf{z}^I_i$ in (\ref{eqn:logistic-reg}) with mean pooling $\frac{1}{k}\sum_{q=1}^k \mathbf{z}^I_{i_q}$. Geometrically speaking, we now optimize the separating hyperplane with respect to the centroids of the positive and negative purchase sequences in item-in embedding space. The sketch in also provided in Figure \ref{fig:geometric2}. Representing sequences by their centroids helps us capturing the \textbf{higher-order} complementariness beyond the pairwise setting.

\subsection{Cold Start Item Inference}
\label{sec:cold-start}

Another advantage of encoding contextual knowledge to item embedding is that we are able to infer item-in embedding for cold-start items using only the trained context embedding $Z^W$. Therefore, the model gives better coverage and saves us the effort of retraining the whole model, especially for large e-commerce platforms, where customers frequently explore only a tiny fraction of products. 

Recall that $\langle \tilde{Z}^I, Z^I \rangle$ measures complementariness and $\langle \tilde{Z}^I, Z^W \rangle$ accounts for contextual relationship. The \textbf{transductive} property is preserved on item-in embedding space because items with similar complementary products and contexts are now embedded closer to each other. Although there may not be enough data to unveil the complementary patterns for cold-start products, we can still embed them by leveraging the contextual relationship. For item $i_0$ with contextual features $\mathbf{w}_{i_0}$, we infer the item-in embedding $\mathbf{z}^I_{i_0}$ according to (\ref{eqn:infer-embedding}) so that it stays closely to contextually similar items in item-in embedding space. As a result, they are more likely to share similar complementary items in recommendation tasks.

\begin{equation}
\label{eqn:infer-embedding}
    \hat{\mathbf{z}} = \arg\max_{\mathbf{z} \in \mathbb{R}^{P}} \prod_{w \in \mathbf{w}_{i_0}} p(w |\mathbf{z}), \text{ where }  p(w |\mathbf{z}) \approx \frac{e^{s(w,i_0)}}{\sum_{\tilde{w} \in \text{Neg}(\mathcal{W})} e^{s(\tilde{w},i_0)}}.
\end{equation}

Inferring item-out embedding for cold-start item, however, is not of pressing demand since the platforms rarely recommend cold-start products to customers. We examine the quality of inferred representations in recommendations tasks in Section \ref{sec:experiment}.

\section{Experiment and Result}
\label{sec:experiment}

We thoroughly evaluate the proposed approach on two tasks: product classification and recommendation using only the optimized product representations. While the classification tasks aim at examining the \textbf{meaningfulness} of product representation, recommendation tasks reveal their \textbf{usefulness} \cite{wan2018representing}. We then examine the representations inferred for cold-start products on the same tasks. Ablation studies are conducted to show the importance of incorporating contextual knowledge and user bias term, and the sensitivity analysis are provided for the embedding dimensions. We present several case studies to show that our approach captures the desired properties of complementariness from Section \ref{sec:Introduction}.

\subsection{Datasets}
\label{sec:data}
We work on two real-world datasets: the public \textsl{Instacart} dataset and a proprietary e-commerce dataset from \textsl{Walmart.com}.

\begin{itemize}
    \item \textbf{\textsl{Instacart.}}
    \textsl{Instacart.com} provides an online service for same-day grocery delivery in the US. The dataset contains around 50 thousand grocery products with shopping records of $\sim$200 thousand users, with a total of $\sim$3 million orders. All products have contextual information, including name, category, department. User contexts are not available. Time information is not presented, but the order of user activities are available.
    
    \item \textbf{Walmart.com.}
    We obtain a proprietary dataset of the\\ Walmart.com whose online shopping catalogue covers more than 100 million items. The dataset consists of user add-to-cart and transaction records (with temporal information) over a specific time span, with $\sim$8 million products, $\sim$20 million customers and a total of $\sim$0.1 billion orders. Product contextual information, including name, category, department and brand are also available. Also, user-segmentation (persona) labels are available for a small fraction of users. 
\end{itemize}

\textbf{Data preprocessing.} We follow the same preprocessing procedure for \textsl{Instacart} dataset described in \cite{wan2018representing}. Items with less than ten transactions are removed. Since temporal information is not available, we use each user's last order as testing data, the second last orders as validation data and the other orders as training data. For the \textsl{Walmart.com} dataset we also filter out items with less than ten transactions. We split the data into training, validation and testing datasets using cutoff times in the chronological order. To construct purchase sequence, we use purchases made in past $d_1$ days as purchase sequence, where $d_1$ can be treated as a tuning parameter that varies for different use cases. For the \textsl{Instacart} dataset we have to make the compromise of using past $k$ purchases since time is not provided. Here $k$ can be treated as sliding window size.

\subsection{Implementation}
\label{sec:implementation}

After observing that the loss function of the proposed approach is separable and sparse, we adopt the Hogwild! \cite{NIPS2011_4390} as optimization algorithm and implement it in C++ for best performance.

Our implementation only takes 15 minutes to train on \textsl{Instacart} dataset and 11 hours to train on the proprietary dataset, which contains $\sim$2 billion observations from the $\sim$5 million users and $\sim$2 million items after filtering, in a Linux server with 16 CPU threads and 40 GB memory. 

\subsection{Results on \textsl{Instacart} dataset}
\label{sec:results_instacart}

We compare the performance of the proposed approach on product classification and with-in basket recommendation against several state-of-the-art product representation learning methods: \textsl{item2vec}, \textsl{prod2vec}, \textsl{metapath2vec} and \textsl{triplet2vec}. We also conduct ablation studies and sensitivity analysis. In ablation studies, we report the performances where item context or user bias term is removed. In the sensitivity analysis, we focus on the dimensions of the item and user embedding. Finally, we evaluate the proposed approach for item cold-start inference. We randomly remove a fraction of items from training dataset as cold-start items and infer their representations after training with the remaining data. The inferred representations are then evaluated using the same two tasks.

\textbf{Tasks and metric}. In product classification tasks we use items' department and categories as labels to evaluate the product representations under coarse-grained and fine-grained scenarios, respectively. We apply the one-vs-all linear logistics regression with the learned item embeddings as input features. The label fraction is 0.5 and $l_2$ regularization term is selected from 5-fold cross-validation. For this imbalanced multi-class classification problem, we report the average \textbf{micro-F1} and \textbf{macro-F1} scores. To prevent information leaks, we exclude department and category information and only use product name and brand for the proposed method.

For the within-basket recommendation task, we rank candidate products according to their complementariness scores to the current basket, by taking the average item-in embeddings as basket representation. The Area Under the ROC Curve (\textbf{AUC}) \cite{rendle2009bpr} and Normalized Discounted Cumulative Gain (\textbf{NDCG}) designed for evaluating recommendation outcome are used as metrics. While \textbf{AUC} examines the average performance without taking account of exact ranking status, \textbf{NDCG} is top-biased, i.e. higher ranked recommendations are given more credit. The classification and recommendations performances are reported in Table \ref{tab:instacart}.

\begin{figure}
    \centering
    \includegraphics[scale=0.25]{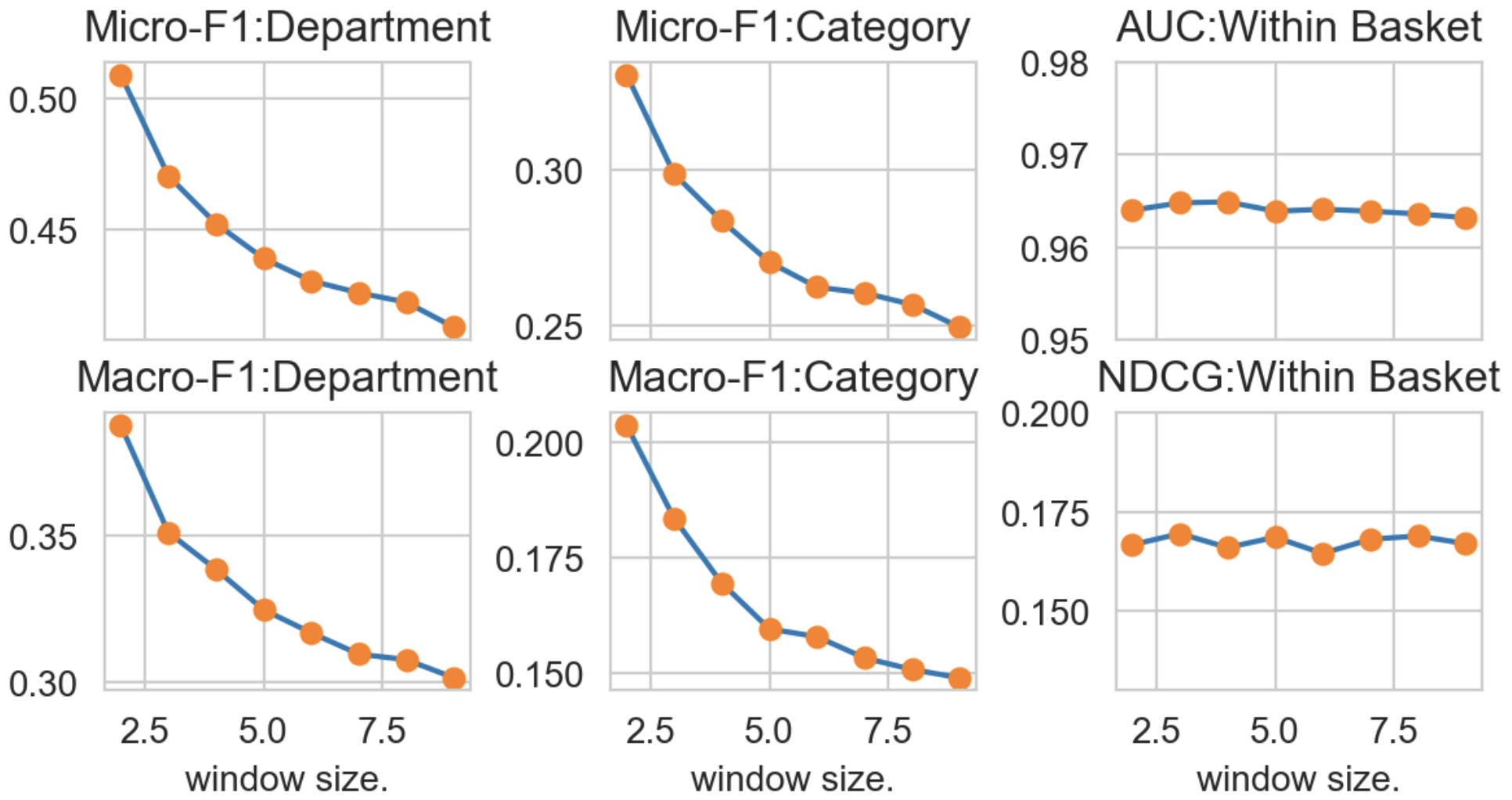}
    \caption{\small The performances of classification and recommendation tasks under different window size $k$ on \textsl{Instacart} dataset without item context. The results are averaged over five runs.}
    \label{fig:ins_wsize}
\end{figure}

First we take a look at how  $k$ (sequence length) influence the performances in both tasks (Figure \ref{fig:ins_wsize}). For fair comparisons with the baseline models reported in \cite{wan2018representing}, we also set product embedding dimension to 32. In Figure \ref{fig:ins_wsize} we see that smaller window size leads to better performances on classification tasks, and for the recommendation task the performances are quite stable under all $k$.  The former result suggests that for grocery shopping on \textsl{Instacart.com}, more recent purchases play major roles in fulfilling the \textbf{meaningfulness} of product representation. However, the latter result indicates that for the recommendation task, our approach is more robust against the length of input sequences, even though longer sequences may introduce more noise. Without further notice, we use $k=2$ in the following experiments and analysis.

From Table \ref{tab:instacart} we observe that the proposed approach (\textbf{full model}) outperforms all baselines, in both classification and recommendation tasks, which suggests that our model can better capture the \textbf{usefulness} and \textbf{meaningfulness} of product representations. By taking out the user bias term, we observe a drop in performances (\textbf{no user} in Table \ref{tab:instacart}) that indicates the importance of explicitly modelling with user preference. Although removing product context leads to worse performance on product classification (\textbf{no context} in Table \ref{tab:instacart}), it gives similar or slightly better results in the recommendation task. We conclude the reasons as follow. Firstly, it is apparent that including context information can improve the \textbf{meaningfulness} of product representation since more useful information is encoded. Secondly, as we discussed before, the impact of context knowledge on recommendation can depend on the sparsity of user-item interactions. Product-context interactions are introduced to deal with the sparsity issue by taking advantage of the \textbf{transductive} property, i.e. similar products may share complementary products.
As a consequence, context information may not provide much help in the denser \textsl{Instacart} dataset which covers only 50 thousand items. We show later that context information is vital for recommendation performances in the sparse proprietary dataset. Finally, the inference algorithm (\ref{eqn:infer-embedding}) of cold-start products gives reasonable results, as they achieve comparable performances with baseline methods in both tasks. It also outperforms the heuristic method which assigns representations to cold start products according to their most similar product (according to the Jaccard similarity of the contexts).

\begin{table}[!htb]
    \centering
    \begin{tabular}{ c c c c c | c c} 
     & \multicolumn{4}{c}{\textbf{Classification}} &\multicolumn{2}{c}{\textbf{Recom.}} \\
     & \multicolumn{2}{c}{\textbf{Department}}& \multicolumn{2}{c}{\textbf{Category}} & \multicolumn{2}{c}{\textbf{In-basket}} \\
     \hline
     \textbf{Method}  & micro & macro & micro & macro & AUC & NDCG \\
     \hline
     
     \textsl{item2vec} & 0.377 & 0.283 & 0.187 & 0.075 & 0.941 & 0.116  \\ 
     \textsl{prod2vec} & 0.330 & 0.218 & 0.106 & 0.030 & 0.941  & 0.125 \\ 
     \textsl{m.2vec$^*$} & 0.331 & 0.221& 0.155 & 0.036 & 0.944 &  0.125 \\ 
     \textsl{triple2vec} & 0.382 & 0.294 & 0.189 & 0.082 & 0.960 & 0.127 \\
     \textbf{no context} & 0.509 & 0.470 & 0.452 & 0.438 & 0.964 & \underline{0.166} \\
     \textbf{no user} & 0.661 & 0.545 & 0.615 & 0.529 & 0.954 & 0.160 \\
     \textbf{full model} & \underline{0.666} & \underline{0.553} & \underline{0.619} & \underline{0.535} & \underline{0.965} & 0.151 \\
     \textbf{cold start} & 0.301 & 0.207 & 0.304 & 0.214 & 0.846 & 0.114 \\
     -infer & - & - & - & - & 0.801 & 0.084 \\
     \hline \\
    \end{tabular}
    \caption{\small Performances on multiple tasks on Instacart data. Methods in bold fonts represents our method under several configurations. The left 4 columns provide micro and macro f-1 scores for the product classification tasks under coarse-grained (department) and fine-grained (category) levels. The right two columns gives the \textbf{AUC} and \textbf{NDCG} for within-basket recommendations. The '-infer' row gives cold start recommendation results based only on context Jaccard similarity without using inference algorithm. $^{*}$\textsl{m.2vec} is the shorthand for \textsl{metapath2vec}. The results are averaged over five runs.} 
    \label{tab:instacart}
\end{table}

\begin{figure}
    \centering
    \includegraphics[scale=0.255]{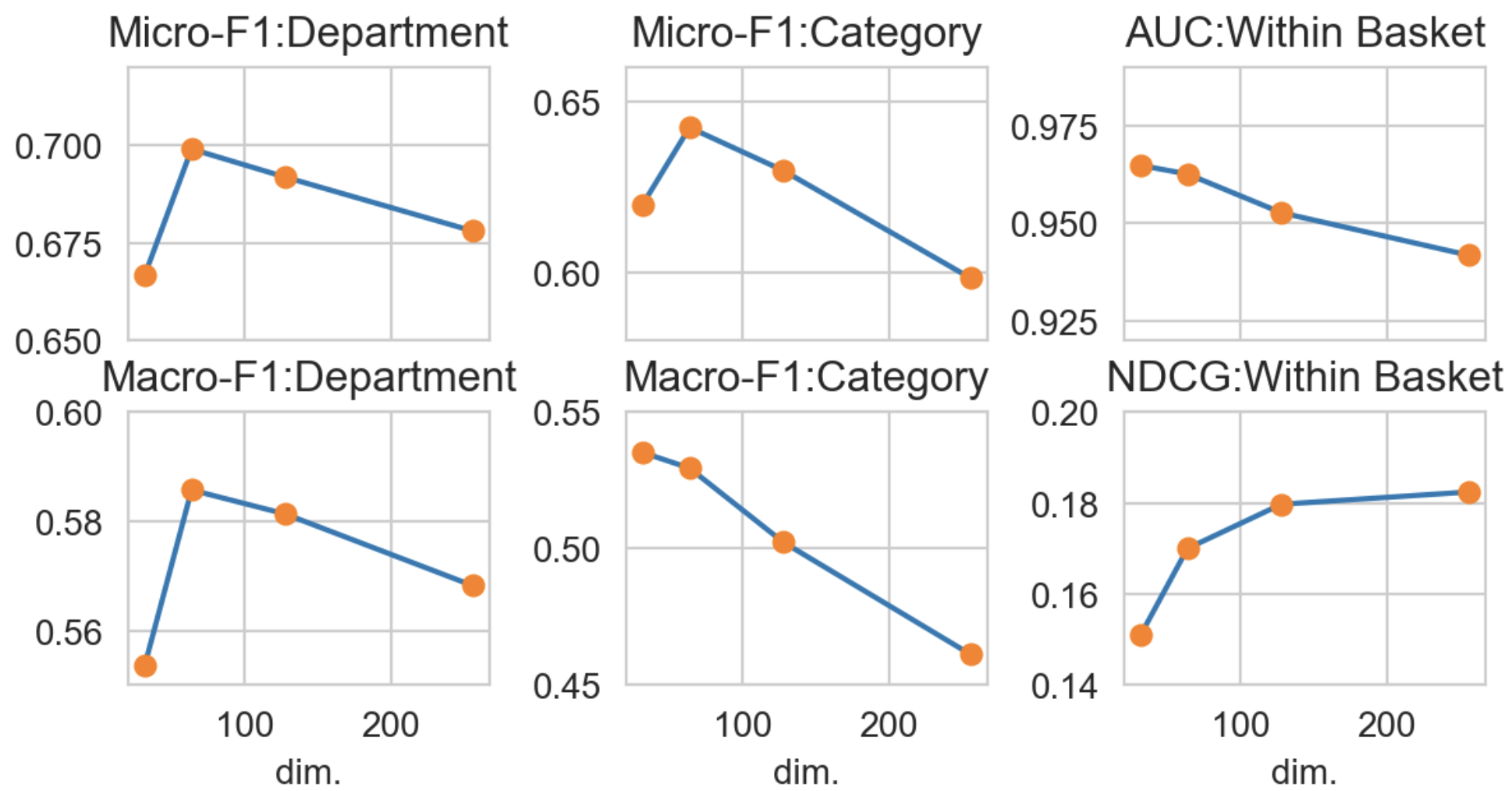}
    \caption{\small Sensitivity analysis on embedding dimension for product classification and within-basket recommendation tasks on \textsl{Instacart} dataset. The results are averaged over five runs.}
    \label{fig:cart_dim}
\end{figure}

\subsection{Results on the \textsl{Walmart.com} dataset}
\label{sec:results_wmt}

We focus on examining the \textbf{usefulness} of our product representation learning method with the \textsl{Walmart.com} dataset. As we mentioned before, the availability of timestamps and the lack of \textbf{confounding} factors such as loyalty make the \textsl{Walmart.com} dataset ideal for evaluating recommendation performances. This also encourages us to compare with baseline models such as collaborative filtering (\textsl{CF}) for recommendation. We also include state-of-the-art supervised implicit-feedback recommendation baselines such as factorizing personalized Markov chain (\textsl{FPMC}) \cite{rendle2010factorizing} and Bayesian personalized ranking (\textsl{BPR}) \cite{rendle2009bpr}. Last but not least, we further compare with the graph-based recommendation method adapted from \textsl{node2vec} \cite{wang2018billion}. On the other hand, without major modifications, the aforementioned product embedding methods do not scale to the size of the proprietary dataset. Still, we manage to implement a similar version of \textsl{prod2vec} based on the \textsl{word2vec} implementation for a complete comparison.

\textbf{Tasks and metric.} The general next-purchase recommendations (includes within-basket and next-basket ) are crucial for real-world recommender systems and are often evaluated by the top-K hitting rate (\textsl{Hit@K}) and top-K normalized discounted cumulative gain (\textsl{NDCG@K}) according to customers' purchase in next $d_2$ days. With timestamps available for the proprietary dataset we can carry out these tasks and metrics. 

\begin{figure}
    \centering
    \includegraphics[scale=0.23]{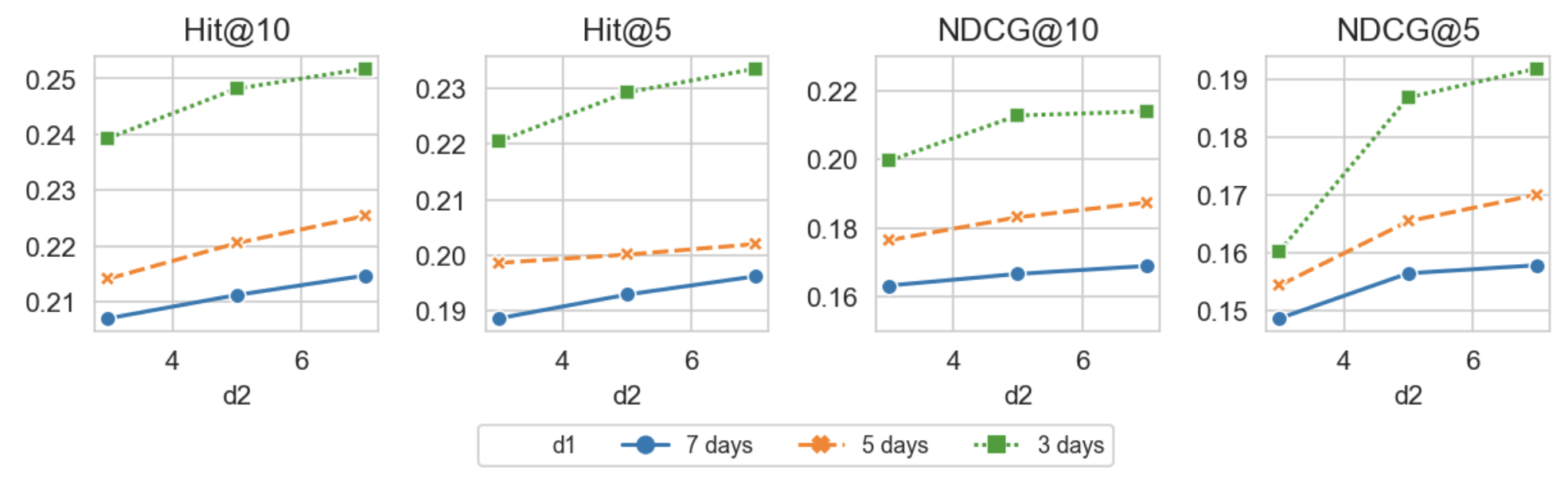}
    \caption{\small Sensitivity analysis for the $d1$ and $d2$, where purchases up to $d1$ days before the current time are used as input sequences and use the user's purchases in next $d2$ days are used as labels, with $P=100$. The results are averaged over five runs.}
    \label{fig:days_sensitivity}
\end{figure}

The performance of the proposed approach under different $d_1$ and $d_2$ are first provided in Figure \ref{fig:days_sensitivity}, where we observe that larger $d_2$ and smaller $d_1$ leads to better outcomes. We notice that different $d_1$ gives very similar results, which is in accordance with the window size in \textsl{Instacart} dataset. They both suggest that our approach is robust against the input sequence length. The fact that larger $d_2$ leads to better results is self-explanatory. Without further notice we use $d_1=3$ to construct input sequences and $d_2=7$ to extract target products for all models. For fair comparisons with baselines which cannot take advantage of user context information, we also exclude user contextual features in our model in all experiments.

\begin{table}[htb]
    \centering
    \begin{tabular}{c c c c c}
    \hline
    Model & Hit@10 & Hit@5 & NDCG@10 & NDCG@5 \\
    \hline
    \textsl{CF} & 0.0962 &  0.0716 & 0.0582 & 0.0506 \\
    \textsl{FPMC} & 0.1473 & 0.1395 & 0.1386 & 0.1241 \\ 
    \textsl{BPR} & 0.1578  & 0.1403 & 0.1392  & 0.1239  \\
    \textsl{node2vec} & 0.1103 & 0.0959  & 0.0874 & 0.0790 \\
    \textsl{prod2vec} & 0.1464 & 0.1321 & 0.1235 & 0.1162 \\
    \textbf{no context}  & 0.1778 &  0.1529 & 0.1346 & 0.1198 \\ 
    \textbf{no user}  & 0.1917 &  0.1726 & 0.1485 & 0.1301 \\ 
    \textbf{full model} & \underline{0.2518} & \underline{0.2336} & \underline{0.2139} & \underline{0.1919} \\
    \textbf{cold start}  & 0.1066 &  0.0873 & 0.0759 & 0.0562 \\
    -infer & 0.0734 & 0.0502 & 0.0531 & 0.0428 \\
    \hline
    
    \end{tabular}
    \caption{\small Comparision with baseline models under $P=100$, $d_1=3$ and $d_2=7$. The results are averaged over five runs.}
    \label{tab:compare}
\end{table}

The recommendation performances are provided in Table \ref{tab:compare}. The proposed approach outperforms all baseline methods, both standard recommendation algorithms and product embedding methods, by significant margins on all metrics. Due to the high sparsity and low signal-to-noise ratio of the Walmart.com dataset, the baseline methods may not be able to capture co-purchase signals without accounting for contextual information and user bias.

As expected, after removing product context terms the performances drop significantly, which again confirms the usefulness of contexts for large sparse datasets. Taking out user bias term also deteriorates the overall performance, which matches what we observe from the \textsl{Instacart} dataset. Besides, the representations inferred for cold-start products give reasonable performances that are comparable to the outcome of some baselines on standard products.

\begin{figure}
    \centering
    \includegraphics[scale=0.25]{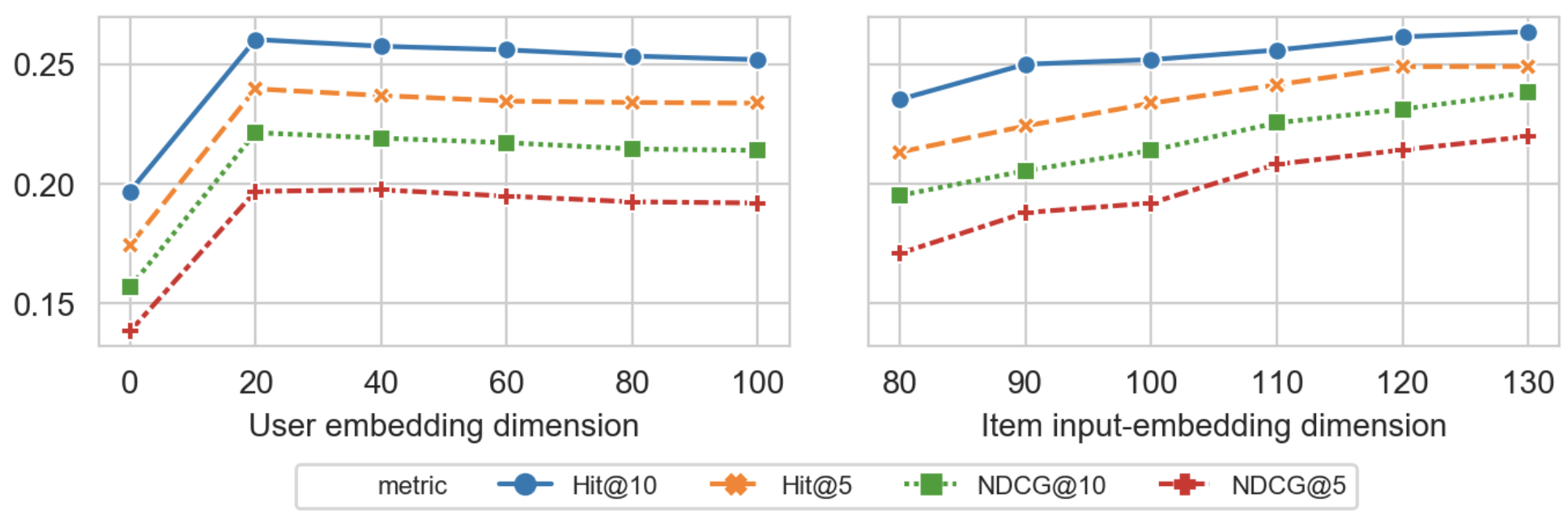}
    \caption{\small Sensitivity analysis for user and item embeddings dimensions on the proprietary dataset averaged over five runs.}
    \label{fig:dim_sensitivity}
\end{figure}

\subsection{Sensitivity analysis}
\label{sec:sensitivity}
We provide a thorough sensitivity analysis on embedding dimensions on both datasets in Figure \ref{fig:cart_dim} and \ref{fig:dim_sensitivity}. Specifically, for \textsl{Instacart} dataset, after increasing product embedding dimension over 80, the proposed model starts to lose classification accuracies. On the other hand, the \textbf{NDCG} for recommendation task keeps improving with larger dimensions (within the range considered), though the \textbf{AUC} slightly decreases. For the Walmart.com dataset, we focus on both user and product embedding dimensions. Increasing user embedding dimension over 20 starts to cause worse performances. Moreover, similar to the results on \textsl{Instacart} dataset, larger product embedding dimensions give better outcomes. 

In fact, recent work reveals the connection between embedding size and the bias-variance trade-off \cite{yin2018dimensionality}. Embedding with large dimensions can overfit training data, and with small dimensions it might underfit. It may help explain our results where a larger product embedding dimension causes overfitting issues in classification tasks on \textsl{Instacart} dataset. For the Walmart.com dataset, increasing user embedding dimension over the threshold may also cause overfitted the user preferences in training data.

\begin{table}[!htb]
\footnotesize
\centering
\begin{tabular}{L{1.8cm}|L{1.8cm} L{1.8cm} L{1.8cm} }
\hline 
    \textbf{Anchor item} &  \multicolumn{3}{c}{\textbf{Top recommended complementary items}} \\
    \hline
    Bath curtain  & Bath curtain hooks & Bath curtain rod & Bath mat \\
    \hline 
    Bath mat & Towels & Mops & Detergent \\
    \hline 
    XBox Player  & Wireless controller & Video Game & Controller stack \\
    \hline 
    TV  & Protection plan & HDMI cable & TV mount frame \\
    \hline
    TV mount frame  & HDMI cable  & DVD hold with cable & Cable cover \\
    \hline
    Round cake pan & Straight spatula & Cupcake pan & Lever tool \\
    \hline 
    Straight spatula & Spatula set & Wood spatula & Steel spatula \\
    \hline
    Round cake pan  & &  & \\
    + & Cake carrier  & Cake decorating set & Icing smoother \\
    Straight spatula & & &\\ 
     \hline
    Sofa & Sofa pillows & Coffee table & Sofa cover \\\hline
    Sofa  & &  & \\
    + & Floor lamp  & Area rug & Sofa pillows \\
    Coffee table & & &\\ \hline
\end{tabular}{}
\caption{\small Case studies for product recommendation.}
\label{tab:case_study}
\end{table}

\subsection{Case study}
\label{sec:cases}

We provide several case studies in Table \ref{tab:case_study} to show that the proposed approach captures the desired properties of complementariness, i.e \textbf{asymmetric}, \textbf{non-transitive} and \textbf{higher-order}. The \textbf{transductive} property has been explicitly modelled by including contextual information, so we do not provide concrete examples here.

By virtue of the examples, we observe the \textbf{Asymmetric} property via \textsl{TV} $\to$ \textsl{TV mount frame} but \textsl{TV mount frame} $\not \to$ \textsl{TV}. The \textbf{Non-transitive} is also reflected by \textsl{TV} $\to$ \textsl{TV mount frame}, \textsl{TV mount frame} $\to$ \textsl{cable cover} but \textsl{TV} $\not \to$ \textsl{cable cover}. As for the \textbf{Higher-order} property, we first note that when \textsl{round cake pan} is combined with \textsl{straight spatula}, the recommendations are not a simple mixture of individual recommendations. \textsl{Cake carrier} and \textsl{cake decorating set}, which complement the combo as a whole, are recommended. Similar higher-order pattern is also observed for \textsl{sofa} and \textsl{coffee table} combination. They provide evidence that our model captures the \textbf{higher-order} complementary patterns.

\section{Conclusion}
\label{sec:conclusion}
In this work, we thoroughly investigate the properties of product complementary relationship: \textbf{asymmetric}, \textbf{non-transitive}, \textbf{transductive} and \textbf{higher-order}. We propose a novel representation learning model for complementary products with dual embeddings that are compatible with the above properties. Contextual information is leveraged to deal with the sparsity in customer-product interactions, and user preferences are explicitly modelled to take account of noise in user purchase sequences. A fast inference algorithm is developed for cold-start products. We demonstrate the effectiveness of the product representations obtained from the proposed approach through classification and recommendation tasks on \textsl{Instacart} and the proprietary e-commerce dataset.

Future work includes exploring populational contexts such as product popularity as well as dynamic factors including trend and seasonality. Furthermore, the idea of context-aware representation learning for complex relationships can be extended beyond e-commerce setting for learning general event representations from user-event interaction sequences.

%
\bibliographystyle{ACM-Reference-Format}
\bibliography{sample-base}


\begin{thebibliography}{40}


\ifx \showCODEN    \undefined \def \showCODEN     #1{\unskip}     \fi
\ifx \showDOI      \undefined \def \showDOI       #1{#1}\fi
\ifx \showISBNx    \undefined \def \showISBNx     #1{\unskip}     \fi
\ifx \showISBNxiii \undefined \def \showISBNxiii  #1{\unskip}     \fi
\ifx \showISSN     \undefined \def \showISSN      #1{\unskip}     \fi
\ifx \showLCCN     \undefined \def \showLCCN      #1{\unskip}     \fi
\ifx \shownote     \undefined \def \shownote      #1{#1}          \fi
\ifx \showarticletitle \undefined \def \showarticletitle #1{#1}   \fi
\ifx \showURL      \undefined \def \showURL       {\relax}        \fi
\providecommand\bibfield[2]{#2}
\providecommand\bibinfo[2]{#2}
\providecommand\natexlab[1]{#1}
\providecommand\showeprint[2][]{arXiv:#2}

\bibitem[\protect\citeauthoryear{Balabanovi{\'c} and Shoham}{Balabanovi{\'c}
  and Shoham}{1997}]%
        {balabanovic1997fab}
\bibfield{author}{\bibinfo{person}{Marko Balabanovi{\'c}} {and}
  \bibinfo{person}{Yoav Shoham}.} \bibinfo{year}{1997}\natexlab{}.
\newblock \showarticletitle{Fab: content-based, collaborative recommendation}.
\newblock \bibinfo{journal}{\emph{Commun. ACM}} \bibinfo{volume}{40},
  \bibinfo{number}{3} (\bibinfo{year}{1997}), \bibinfo{pages}{66--72}.
\newblock


\bibitem[\protect\citeauthoryear{Barkan and Koenigstein}{Barkan and
  Koenigstein}{2016}]%
        {barkan2016item2vec}
\bibfield{author}{\bibinfo{person}{Oren Barkan} {and} \bibinfo{person}{Noam
  Koenigstein}.} \bibinfo{year}{2016}\natexlab{}.
\newblock \showarticletitle{Item2vec: neural item embedding for collaborative
  filtering}. In \bibinfo{booktitle}{\emph{2016 IEEE 26th International
  Workshop on Machine Learning for Signal Processing (MLSP)}}. IEEE,
  \bibinfo{pages}{1--6}.
\newblock


\bibitem[\protect\citeauthoryear{Belkin and Niyogi}{Belkin and Niyogi}{2003}]%
        {belkin2003laplacian}
\bibfield{author}{\bibinfo{person}{Mikhail Belkin} {and}
  \bibinfo{person}{Partha Niyogi}.} \bibinfo{year}{2003}\natexlab{}.
\newblock \showarticletitle{Laplacian eigenmaps for dimensionality reduction
  and data representation}.
\newblock \bibinfo{journal}{\emph{Neural computation}} \bibinfo{volume}{15},
  \bibinfo{number}{6} (\bibinfo{year}{2003}), \bibinfo{pages}{1373--1396}.
\newblock


\bibitem[\protect\citeauthoryear{Bobadilla, Ortega, Hernando, and
  Bernal}{Bobadilla et~al\mbox{.}}{2012}]%
        {bobadilla2012collaborative}
\bibfield{author}{\bibinfo{person}{Jes{\'u}S Bobadilla},
  \bibinfo{person}{Fernando Ortega}, \bibinfo{person}{Antonio Hernando}, {and}
  \bibinfo{person}{Jes{\'u}S Bernal}.} \bibinfo{year}{2012}\natexlab{}.
\newblock \showarticletitle{A collaborative filtering approach to mitigate the
  new user cold start problem}.
\newblock \bibinfo{journal}{\emph{Knowledge-Based Systems}}
  \bibinfo{volume}{26} (\bibinfo{year}{2012}), \bibinfo{pages}{225--238}.
\newblock


\bibitem[\protect\citeauthoryear{Bojanowski, Grave, Joulin, and
  Mikolov}{Bojanowski et~al\mbox{.}}{2017}]%
        {bojanowski2017enriching}
\bibfield{author}{\bibinfo{person}{Piotr Bojanowski}, \bibinfo{person}{Edouard
  Grave}, \bibinfo{person}{Armand Joulin}, {and} \bibinfo{person}{Tomas
  Mikolov}.} \bibinfo{year}{2017}\natexlab{}.
\newblock \showarticletitle{Enriching word vectors with subword information}.
\newblock \bibinfo{journal}{\emph{Transactions of the Association for
  Computational Linguistics}}  \bibinfo{volume}{5} (\bibinfo{year}{2017}),
  \bibinfo{pages}{135--146}.
\newblock


\bibitem[\protect\citeauthoryear{Covington, Adams, and Sargin}{Covington
  et~al\mbox{.}}{2016}]%
        {covington2016deep}
\bibfield{author}{\bibinfo{person}{Paul Covington}, \bibinfo{person}{Jay
  Adams}, {and} \bibinfo{person}{Emre Sargin}.}
  \bibinfo{year}{2016}\natexlab{}.
\newblock \showarticletitle{Deep neural networks for youtube recommendations}.
  In \bibinfo{booktitle}{\emph{Proceedings of the 10th ACM Conference on
  Recommender Systems}}. ACM, \bibinfo{pages}{191--198}.
\newblock


\bibitem[\protect\citeauthoryear{Dong, Chawla, and Swami}{Dong
  et~al\mbox{.}}{2017}]%
        {dong2017metapath2vec}
\bibfield{author}{\bibinfo{person}{Yuxiao Dong}, \bibinfo{person}{Nitesh~V
  Chawla}, {and} \bibinfo{person}{Ananthram Swami}.}
  \bibinfo{year}{2017}\natexlab{}.
\newblock \showarticletitle{metapath2vec: Scalable representation learning for
  heterogeneous networks}. In \bibinfo{booktitle}{\emph{Proceedings of the 23rd
  ACM SIGKDD international conference on knowledge discovery and data mining}}.
  ACM, \bibinfo{pages}{135--144}.
\newblock


\bibitem[\protect\citeauthoryear{Gantner, Drumond, Freudenthaler, Rendle, and
  Schmidt-Thieme}{Gantner et~al\mbox{.}}{2010}]%
        {gantner2010learning}
\bibfield{author}{\bibinfo{person}{Zeno Gantner}, \bibinfo{person}{Lucas
  Drumond}, \bibinfo{person}{Christoph Freudenthaler}, \bibinfo{person}{Steffen
  Rendle}, {and} \bibinfo{person}{Lars Schmidt-Thieme}.}
  \bibinfo{year}{2010}\natexlab{}.
\newblock \showarticletitle{Learning attribute-to-feature mappings for
  cold-start recommendations}. In \bibinfo{booktitle}{\emph{Data Mining (ICDM),
  2010 IEEE 10th International Conference on}}. IEEE,
  \bibinfo{pages}{176--185}.
\newblock


\bibitem[\protect\citeauthoryear{Grbovic, Radosavljevic, Djuric, Bhamidipati,
  Savla, Bhagwan, and Sharp}{Grbovic et~al\mbox{.}}{2015}]%
        {grbovic2015commerce}
\bibfield{author}{\bibinfo{person}{Mihajlo Grbovic}, \bibinfo{person}{Vladan
  Radosavljevic}, \bibinfo{person}{Nemanja Djuric}, \bibinfo{person}{Narayan
  Bhamidipati}, \bibinfo{person}{Jaikit Savla}, \bibinfo{person}{Varun
  Bhagwan}, {and} \bibinfo{person}{Doug Sharp}.}
  \bibinfo{year}{2015}\natexlab{}.
\newblock \showarticletitle{E-commerce in your inbox: Product recommendations
  at scale}. In \bibinfo{booktitle}{\emph{Proceedings of the 21th ACM SIGKDD
  International Conference on Knowledge Discovery and Data Mining}}. ACM,
  \bibinfo{pages}{1809--1818}.
\newblock


\bibitem[\protect\citeauthoryear{Grover and Leskovec}{Grover and
  Leskovec}{2016}]%
        {grover2016node2vec}
\bibfield{author}{\bibinfo{person}{Aditya Grover} {and} \bibinfo{person}{Jure
  Leskovec}.} \bibinfo{year}{2016}\natexlab{}.
\newblock \showarticletitle{node2vec: Scalable feature learning for networks}.
  In \bibinfo{booktitle}{\emph{Proceedings of the 22nd ACM SIGKDD international
  conference on Knowledge discovery and data mining}}. ACM,
  \bibinfo{pages}{855--864}.
\newblock


\bibitem[\protect\citeauthoryear{Hannon, Bennett, and Smyth}{Hannon
  et~al\mbox{.}}{2010}]%
        {hannon2010recommending}
\bibfield{author}{\bibinfo{person}{John Hannon}, \bibinfo{person}{Mike
  Bennett}, {and} \bibinfo{person}{Barry Smyth}.}
  \bibinfo{year}{2010}\natexlab{}.
\newblock \showarticletitle{Recommending twitter users to follow using content
  and collaborative filtering approaches}. In
  \bibinfo{booktitle}{\emph{Proceedings of the fourth ACM conference on
  Recommender systems}}. ACM, \bibinfo{pages}{199--206}.
\newblock


\bibitem[\protect\citeauthoryear{Huang, Chen, and Zeng}{Huang
  et~al\mbox{.}}{2004}]%
        {huang2004applying}
\bibfield{author}{\bibinfo{person}{Zan Huang}, \bibinfo{person}{Hsinchun Chen},
  {and} \bibinfo{person}{Daniel Zeng}.} \bibinfo{year}{2004}\natexlab{}.
\newblock \showarticletitle{Applying associative retrieval techniques to
  alleviate the sparsity problem in collaborative filtering}.
\newblock \bibinfo{journal}{\emph{ACM Transactions on Information Systems
  (TOIS)}} \bibinfo{volume}{22}, \bibinfo{number}{1} (\bibinfo{year}{2004}),
  \bibinfo{pages}{116--142}.
\newblock


\bibitem[\protect\citeauthoryear{Konstas, Stathopoulos, and Jose}{Konstas
  et~al\mbox{.}}{2009}]%
        {konstas2009social}
\bibfield{author}{\bibinfo{person}{Ioannis Konstas}, \bibinfo{person}{Vassilios
  Stathopoulos}, {and} \bibinfo{person}{Joemon~M Jose}.}
  \bibinfo{year}{2009}\natexlab{}.
\newblock \showarticletitle{On social networks and collaborative
  recommendation}. In \bibinfo{booktitle}{\emph{Proceedings of the 32nd
  international ACM SIGIR conference on Research and development in information
  retrieval}}. ACM, \bibinfo{pages}{195--202}.
\newblock


\bibitem[\protect\citeauthoryear{Koren and Bell}{Koren and Bell}{2015}]%
        {koren2015advances}
\bibfield{author}{\bibinfo{person}{Yehuda Koren} {and} \bibinfo{person}{Robert
  Bell}.} \bibinfo{year}{2015}\natexlab{}.
\newblock \showarticletitle{Advances in collaborative filtering}.
\newblock In \bibinfo{booktitle}{\emph{Recommender systems handbook}}.
  \bibinfo{publisher}{Springer}, \bibinfo{pages}{77--118}.
\newblock


\bibitem[\protect\citeauthoryear{Lam, Vu, Le, and Duong}{Lam
  et~al\mbox{.}}{2008}]%
        {lam2008addressing}
\bibfield{author}{\bibinfo{person}{Xuan~Nhat Lam}, \bibinfo{person}{Thuc Vu},
  \bibinfo{person}{Trong~Duc Le}, {and} \bibinfo{person}{Anh~Duc Duong}.}
  \bibinfo{year}{2008}\natexlab{}.
\newblock \showarticletitle{Addressing cold-start problem in recommendation
  systems}. In \bibinfo{booktitle}{\emph{Proceedings of the 2nd international
  conference on Ubiquitous information management and communication}}. ACM,
  \bibinfo{pages}{208--211}.
\newblock


\bibitem[\protect\citeauthoryear{Le and Mikolov}{Le and Mikolov}{2014}]%
        {le2014distributed}
\bibfield{author}{\bibinfo{person}{Quoc Le} {and} \bibinfo{person}{Tomas
  Mikolov}.} \bibinfo{year}{2014}\natexlab{}.
\newblock \showarticletitle{Distributed representations of sentences and
  documents}. In \bibinfo{booktitle}{\emph{International Conference on Machine
  Learning}}. \bibinfo{pages}{1188--1196}.
\newblock


\bibitem[\protect\citeauthoryear{Linden, Smith, and York}{Linden
  et~al\mbox{.}}{2003}]%
        {linden2003amazon}
\bibfield{author}{\bibinfo{person}{Greg Linden}, \bibinfo{person}{Brent Smith},
  {and} \bibinfo{person}{Jeremy York}.} \bibinfo{year}{2003}\natexlab{}.
\newblock \showarticletitle{Amazon. com recommendations: Item-to-item
  collaborative filtering}.
\newblock \bibinfo{journal}{\emph{IEEE Internet computing}} \bibinfo{number}{1}
  (\bibinfo{year}{2003}), \bibinfo{pages}{76--80}.
\newblock


\bibitem[\protect\citeauthoryear{McAuley, Pandey, and Leskovec}{McAuley
  et~al\mbox{.}}{2015}]%
        {mcauley2015inferring}
\bibfield{author}{\bibinfo{person}{Julian McAuley}, \bibinfo{person}{Rahul
  Pandey}, {and} \bibinfo{person}{Jure Leskovec}.}
  \bibinfo{year}{2015}\natexlab{}.
\newblock \showarticletitle{Inferring networks of substitutable and
  complementary products}. In \bibinfo{booktitle}{\emph{Proceedings of the 21th
  ACM SIGKDD International Conference on Knowledge Discovery and Data Mining}}.
  ACM, \bibinfo{pages}{785--794}.
\newblock


\bibitem[\protect\citeauthoryear{Melville, Mooney, and Nagarajan}{Melville
  et~al\mbox{.}}{2002}]%
        {melville2002content}
\bibfield{author}{\bibinfo{person}{Prem Melville}, \bibinfo{person}{Raymond~J
  Mooney}, {and} \bibinfo{person}{Ramadass Nagarajan}.}
  \bibinfo{year}{2002}\natexlab{}.
\newblock \showarticletitle{Content-boosted collaborative filtering for
  improved recommendations}.
\newblock \bibinfo{journal}{\emph{Aaai/iaai}}  \bibinfo{volume}{23}
  (\bibinfo{year}{2002}), \bibinfo{pages}{187--192}.
\newblock


\bibitem[\protect\citeauthoryear{Mikolov, Sutskever, Chen, Corrado, and
  Dean}{Mikolov et~al\mbox{.}}{2013}]%
        {mikolov2013distributed}
\bibfield{author}{\bibinfo{person}{Tomas Mikolov}, \bibinfo{person}{Ilya
  Sutskever}, \bibinfo{person}{Kai Chen}, \bibinfo{person}{Greg~S Corrado},
  {and} \bibinfo{person}{Jeff Dean}.} \bibinfo{year}{2013}\natexlab{}.
\newblock \showarticletitle{Distributed representations of words and phrases
  and their compositionality}. In \bibinfo{booktitle}{\emph{Advances in neural
  information processing systems}}. \bibinfo{pages}{3111--3119}.
\newblock


\bibitem[\protect\citeauthoryear{Ou, Cui, Pei, Zhang, and Zhu}{Ou
  et~al\mbox{.}}{2016}]%
        {ou2016asymmetric}
\bibfield{author}{\bibinfo{person}{Mingdong Ou}, \bibinfo{person}{Peng Cui},
  \bibinfo{person}{Jian Pei}, \bibinfo{person}{Ziwei Zhang}, {and}
  \bibinfo{person}{Wenwu Zhu}.} \bibinfo{year}{2016}\natexlab{}.
\newblock \showarticletitle{Asymmetric transitivity preserving graph
  embedding}. In \bibinfo{booktitle}{\emph{Proceedings of the 22nd ACM SIGKDD
  international conference on Knowledge discovery and data mining}}. ACM,
  \bibinfo{pages}{1105--1114}.
\newblock


\bibitem[\protect\citeauthoryear{Papagelis, Plexousakis, and
  Kutsuras}{Papagelis et~al\mbox{.}}{2005}]%
        {papagelis2005alleviating}
\bibfield{author}{\bibinfo{person}{Manos Papagelis}, \bibinfo{person}{Dimitris
  Plexousakis}, {and} \bibinfo{person}{Themistoklis Kutsuras}.}
  \bibinfo{year}{2005}\natexlab{}.
\newblock \showarticletitle{Alleviating the sparsity problem of collaborative
  filtering using trust inferences}.
\newblock In \bibinfo{booktitle}{\emph{Trust management}}.
  \bibinfo{publisher}{Springer}, \bibinfo{pages}{224--239}.
\newblock


\bibitem[\protect\citeauthoryear{Pennington, Socher, and Manning}{Pennington
  et~al\mbox{.}}{2014}]%
        {pennington2014glove}
\bibfield{author}{\bibinfo{person}{Jeffrey Pennington},
  \bibinfo{person}{Richard Socher}, {and} \bibinfo{person}{Christopher
  Manning}.} \bibinfo{year}{2014}\natexlab{}.
\newblock \showarticletitle{Glove: Global vectors for word representation}. In
  \bibinfo{booktitle}{\emph{Proceedings of the 2014 conference on empirical
  methods in natural language processing (EMNLP)}}.
  \bibinfo{pages}{1532--1543}.
\newblock


\bibitem[\protect\citeauthoryear{Recht, Re, Wright, and Niu}{Recht
  et~al\mbox{.}}{2011}]%
        {NIPS2011_4390}
\bibfield{author}{\bibinfo{person}{Benjamin Recht},
  \bibinfo{person}{Christopher Re}, \bibinfo{person}{Stephen Wright}, {and}
  \bibinfo{person}{Feng Niu}.} \bibinfo{year}{2011}\natexlab{}.
\newblock \showarticletitle{Hogwild: A Lock-Free Approach to Parallelizing
  Stochastic Gradient Descent}.
\newblock In \bibinfo{booktitle}{\emph{Advances in Neural Information
  Processing Systems 24}}, \bibfield{editor}{\bibinfo{person}{J.~Shawe-Taylor},
  \bibinfo{person}{R.~S. Zemel}, \bibinfo{person}{P.~L. Bartlett},
  \bibinfo{person}{F.~Pereira}, {and} \bibinfo{person}{K.~Q. Weinberger}}
  (Eds.). \bibinfo{publisher}{Curran Associates, Inc.},
  \bibinfo{pages}{693--701}.
\newblock
\urldef\tempurl%
\url{http://papers.nips.cc/paper/4390-hogwild-a-lock-free-approach-to-parallelizing-stochastic-gradient-descent.pdf}
\showURL{%
\tempurl}


\bibitem[\protect\citeauthoryear{Rendle, Freudenthaler, Gantner, and
  Schmidt-Thieme}{Rendle et~al\mbox{.}}{2009}]%
        {rendle2009bpr}
\bibfield{author}{\bibinfo{person}{Steffen Rendle}, \bibinfo{person}{Christoph
  Freudenthaler}, \bibinfo{person}{Zeno Gantner}, {and} \bibinfo{person}{Lars
  Schmidt-Thieme}.} \bibinfo{year}{2009}\natexlab{}.
\newblock \showarticletitle{BPR: Bayesian personalized ranking from implicit
  feedback}. In \bibinfo{booktitle}{\emph{Proceedings of the twenty-fifth
  conference on uncertainty in artificial intelligence}}. AUAI Press,
  \bibinfo{pages}{452--461}.
\newblock


\bibitem[\protect\citeauthoryear{Rendle, Freudenthaler, and
  Schmidt-Thieme}{Rendle et~al\mbox{.}}{2010}]%
        {rendle2010factorizing}
\bibfield{author}{\bibinfo{person}{Steffen Rendle}, \bibinfo{person}{Christoph
  Freudenthaler}, {and} \bibinfo{person}{Lars Schmidt-Thieme}.}
  \bibinfo{year}{2010}\natexlab{}.
\newblock \showarticletitle{Factorizing personalized markov chains for
  next-basket recommendation}. In \bibinfo{booktitle}{\emph{Proceedings of the
  19th international conference on World wide web}}. ACM,
  \bibinfo{pages}{811--820}.
\newblock


\bibitem[\protect\citeauthoryear{Sarwar, Karypis, Konstan, and Riedl}{Sarwar
  et~al\mbox{.}}{2001}]%
        {sarwar2001item}
\bibfield{author}{\bibinfo{person}{Badrul Sarwar}, \bibinfo{person}{George
  Karypis}, \bibinfo{person}{Joseph Konstan}, {and} \bibinfo{person}{John
  Riedl}.} \bibinfo{year}{2001}\natexlab{}.
\newblock \showarticletitle{Item-based collaborative filtering recommendation
  algorithms}. In \bibinfo{booktitle}{\emph{Proceedings of the 10th
  international conference on World Wide Web}}. ACM, \bibinfo{pages}{285--295}.
\newblock


\bibitem[\protect\citeauthoryear{Saveski and Mantrach}{Saveski and
  Mantrach}{2014}]%
        {saveski2014item}
\bibfield{author}{\bibinfo{person}{Martin Saveski} {and} \bibinfo{person}{Amin
  Mantrach}.} \bibinfo{year}{2014}\natexlab{}.
\newblock \showarticletitle{Item cold-start recommendations: learning local
  collective embeddings}. In \bibinfo{booktitle}{\emph{Proceedings of the 8th
  ACM Conference on Recommender systems}}. ACM, \bibinfo{pages}{89--96}.
\newblock


\bibitem[\protect\citeauthoryear{Schein, Popescul, Ungar, and Pennock}{Schein
  et~al\mbox{.}}{2002}]%
        {schein2002methods}
\bibfield{author}{\bibinfo{person}{Andrew~I Schein},
  \bibinfo{person}{Alexandrin Popescul}, \bibinfo{person}{Lyle~H Ungar}, {and}
  \bibinfo{person}{David~M Pennock}.} \bibinfo{year}{2002}\natexlab{}.
\newblock \showarticletitle{Methods and metrics for cold-start
  recommendations}. In \bibinfo{booktitle}{\emph{Proceedings of the 25th annual
  international ACM SIGIR conference on Research and development in information
  retrieval}}. ACM, \bibinfo{pages}{253--260}.
\newblock


\bibitem[\protect\citeauthoryear{Soboroff and Nicholas}{Soboroff and
  Nicholas}{1999}]%
        {soboroff1999combining}
\bibfield{author}{\bibinfo{person}{Ian Soboroff} {and} \bibinfo{person}{Charles
  Nicholas}.} \bibinfo{year}{1999}\natexlab{}.
\newblock \showarticletitle{Combining content and collaboration in text
  filtering}. In \bibinfo{booktitle}{\emph{Proceedings of the IJCAI}},
  Vol.~\bibinfo{volume}{99}. sn, \bibinfo{pages}{86--91}.
\newblock


\bibitem[\protect\citeauthoryear{Vasile, Smirnova, and Conneau}{Vasile
  et~al\mbox{.}}{2016}]%
        {vasile2016meta}
\bibfield{author}{\bibinfo{person}{Flavian Vasile}, \bibinfo{person}{Elena
  Smirnova}, {and} \bibinfo{person}{Alexis Conneau}.}
  \bibinfo{year}{2016}\natexlab{}.
\newblock \showarticletitle{Meta-prod2vec: Product embeddings using
  side-information for recommendation}. In
  \bibinfo{booktitle}{\emph{Proceedings of the 10th ACM Conference on
  Recommender Systems}}. ACM, \bibinfo{pages}{225--232}.
\newblock


\bibitem[\protect\citeauthoryear{Wainwright, Jordan, et~al\mbox{.}}{Wainwright
  et~al\mbox{.}}{2008}]%
        {wainwright2008graphical}
\bibfield{author}{\bibinfo{person}{Martin~J Wainwright},
  \bibinfo{person}{Michael~I Jordan}, {et~al\mbox{.}}}
  \bibinfo{year}{2008}\natexlab{}.
\newblock \showarticletitle{Graphical models, exponential families, and
  variational inference}.
\newblock \bibinfo{journal}{\emph{Foundations and Trends{\textregistered} in
  Machine Learning}} \bibinfo{volume}{1}, \bibinfo{number}{1--2}
  (\bibinfo{year}{2008}), \bibinfo{pages}{1--305}.
\newblock


\bibitem[\protect\citeauthoryear{Wan, Wang, Liu, Bennett, and McAuley}{Wan
  et~al\mbox{.}}{2018}]%
        {wan2018representing}
\bibfield{author}{\bibinfo{person}{Mengting Wan}, \bibinfo{person}{Di Wang},
  \bibinfo{person}{Jie Liu}, \bibinfo{person}{Paul Bennett}, {and}
  \bibinfo{person}{Julian McAuley}.} \bibinfo{year}{2018}\natexlab{}.
\newblock \showarticletitle{Representing and Recommending Shopping Baskets with
  Complementarity, Compatibility and Loyalty}. In
  \bibinfo{booktitle}{\emph{Proceedings of the 27th ACM International
  Conference on Information and Knowledge Management}}. ACM,
  \bibinfo{pages}{1133--1142}.
\newblock


\bibitem[\protect\citeauthoryear{Wang, Huang, Zhao, Zhang, Zhao, and Lee}{Wang
  et~al\mbox{.}}{2018}]%
        {wang2018billion}
\bibfield{author}{\bibinfo{person}{Jizhe Wang}, \bibinfo{person}{Pipei Huang},
  \bibinfo{person}{Huan Zhao}, \bibinfo{person}{Zhibo Zhang},
  \bibinfo{person}{Binqiang Zhao}, {and} \bibinfo{person}{Dik~Lun Lee}.}
  \bibinfo{year}{2018}\natexlab{}.
\newblock \showarticletitle{Billion-scale Commodity Embedding for E-commerce
  Recommendation in Alibaba}.
\newblock \bibinfo{journal}{\emph{arXiv preprint arXiv:1803.02349}}
  (\bibinfo{year}{2018}).
\newblock


\bibitem[\protect\citeauthoryear{Yin and Shen}{Yin and Shen}{2018}]%
        {yin2018dimensionality}
\bibfield{author}{\bibinfo{person}{Zi Yin} {and} \bibinfo{person}{Yuanyuan
  Shen}.} \bibinfo{year}{2018}\natexlab{}.
\newblock \showarticletitle{On the dimensionality of word embedding}. In
  \bibinfo{booktitle}{\emph{Advances in Neural Information Processing
  Systems}}. \bibinfo{pages}{895--906}.
\newblock


\bibitem[\protect\citeauthoryear{Ying, He, Chen, Eksombatchai, Hamilton, and
  Leskovec}{Ying et~al\mbox{.}}{2018}]%
        {ying2018graph}
\bibfield{author}{\bibinfo{person}{Rex Ying}, \bibinfo{person}{Ruining He},
  \bibinfo{person}{Kaifeng Chen}, \bibinfo{person}{Pong Eksombatchai},
  \bibinfo{person}{William~L Hamilton}, {and} \bibinfo{person}{Jure Leskovec}.}
  \bibinfo{year}{2018}\natexlab{}.
\newblock \showarticletitle{Graph Convolutional Neural Networks for Web-Scale
  Recommender Systems}.
\newblock \bibinfo{journal}{\emph{arXiv preprint arXiv:1806.01973}}
  (\bibinfo{year}{2018}).
\newblock


\bibitem[\protect\citeauthoryear{Zhang, Lu, Niu, and Caverlee}{Zhang
  et~al\mbox{.}}{2018}]%
        {Zhang:2018:QNC:3240323.3240368}
\bibfield{author}{\bibinfo{person}{Yin Zhang}, \bibinfo{person}{Haokai Lu},
  \bibinfo{person}{Wei Niu}, {and} \bibinfo{person}{James Caverlee}.}
  \bibinfo{year}{2018}\natexlab{}.
\newblock \showarticletitle{Quality-aware Neural Complementary Item
  Recommendation}. In \bibinfo{booktitle}{\emph{Proceedings of the 12th ACM
  Conference on Recommender Systems}} \emph{(\bibinfo{series}{RecSys '18})}.
  \bibinfo{publisher}{ACM}, \bibinfo{address}{New York, NY, USA},
  \bibinfo{pages}{77--85}.
\newblock
\showISBNx{978-1-4503-5901-6}
\urldef\tempurl%
\url{https://doi.org/10.1145/3240323.3240368}
\showDOI{\tempurl}


\bibitem[\protect\citeauthoryear{Zheng, Wu, Niu, and Bolivar}{Zheng
  et~al\mbox{.}}{2009}]%
        {zheng2009substitutes}
\bibfield{author}{\bibinfo{person}{Jiaqian Zheng}, \bibinfo{person}{Xiaoyuan
  Wu}, \bibinfo{person}{Junyu Niu}, {and} \bibinfo{person}{Alvaro Bolivar}.}
  \bibinfo{year}{2009}\natexlab{}.
\newblock \showarticletitle{Substitutes or complements: another step forward in
  recommendations}. In \bibinfo{booktitle}{\emph{Proceedings of the 10th ACM
  conference on Electronic commerce}}. ACM, \bibinfo{pages}{139--146}.
\newblock


\bibitem[\protect\citeauthoryear{Zhou, Zhu, Song, Fan, Zhu, Ma, Yan, Jin, Li,
  and Gai}{Zhou et~al\mbox{.}}{2018}]%
        {zhou2018deep}
\bibfield{author}{\bibinfo{person}{Guorui Zhou}, \bibinfo{person}{Xiaoqiang
  Zhu}, \bibinfo{person}{Chenru Song}, \bibinfo{person}{Ying Fan},
  \bibinfo{person}{Han Zhu}, \bibinfo{person}{Xiao Ma},
  \bibinfo{person}{Yanghui Yan}, \bibinfo{person}{Junqi Jin},
  \bibinfo{person}{Han Li}, {and} \bibinfo{person}{Kun Gai}.}
  \bibinfo{year}{2018}\natexlab{}.
\newblock \showarticletitle{Deep interest network for click-through rate
  prediction}. In \bibinfo{booktitle}{\emph{Proceedings of the 24th ACM SIGKDD
  International Conference on Knowledge Discovery \& Data Mining}}. ACM,
  \bibinfo{pages}{1059--1068}.
\newblock


\bibitem[\protect\citeauthoryear{Zhou, Yang, and Zha}{Zhou
  et~al\mbox{.}}{2011}]%
        {zhou2011functional}
\bibfield{author}{\bibinfo{person}{Ke Zhou}, \bibinfo{person}{Shuang-Hong
  Yang}, {and} \bibinfo{person}{Hongyuan Zha}.}
  \bibinfo{year}{2011}\natexlab{}.
\newblock \showarticletitle{Functional matrix factorizations for cold-start
  recommendation}. In \bibinfo{booktitle}{\emph{Proceedings of the 34th
  international ACM SIGIR conference on Research and development in Information
  Retrieval}}. ACM, \bibinfo{pages}{315--324}.
\newblock


\end{thebibliography}

\appendix 
\section{Research Methods}
\label{appen:implementation}

Since we optimize the embeddings with stochastic gradient descent, we first present the gradients for the embeddings with respect to the loss function on single complete observation \\ $\{ u, i_{t}, i_{t-1}, \dots, i_{t-k}, \mathbf{x}_u, \mathbf{w}_{i_t}, \mathbf{w}_{i_{t-1}}, \dots, \mathbf{w}_{i_{t-k}} \}$ presented in (\ref{eqn:loss}) under negative sampling. 

Let $Q_{i,u,i_{t-1:t-k}} = s(i,u) + s(i,i_{t-1:t-k})$, $Q_{w,i} = s(w,i)$ and $Q_{x,u} = s(x,u)$. The gradient for $\mathbf{z}^U_u$ can be computed by
\begin{equation}
\label{eqn:appen1}
\begin{split}
    \nabla_{\mathbf{z}^U_u} \ell =& -\mathbf{z}^I_{i_t} \frac{e^{-Q_{i,u,i_{t-1:t-k}}}}{1 + e^{-Q_{i,u,i_{t-1:t-k}}}}  + \sum_{\tilde{i} \in \text{Neg}(\mathcal{I})} \mathbf{z}^I_{\tilde{i}} \frac{e^{Q_{\tilde{i},u,i_{t-1:t-k}}}}{1 + e^{Q_{\tilde{i},u,i_{t-1:t-k}}}} \\
    &- \sum_{x \in \mathbf{x}_u} \mathbf{z}^X_{x} \frac{e^{-Q_{x,u}}}{1+e^{-Q_{x,u}}} + \sum_{\tilde{x} \in \mathcal{X} \setminus x} \mathbf{z}^X_{\tilde{x}} \frac{e^{Q_{\tilde{x}, u}}}{1 + e^{Q_{\tilde{x}, u}}}.
\end{split}
\end{equation}

The gradient for $\tilde{\mathbf{z}}^I_{i_t}$ and $\mathbf{z}^I_{i_t}$ are given by
\begin{equation}
\label{eqn:appenw}
\begin{split}
    \nabla_{\tilde{\mathbf{z}}^I_{i_t}} \ell =& -\Big(\mathbf{z}^U_{u} - \frac{1}{k}\sum_{j=1}^k \mathbf{z}^I_{i_{t-j}}\Big) \frac{e^{-Q_{i,u,i_{t-1:t-k}}}}{1 + e^{-Q_{i,u,i_{t-1:t-k}}}},
\end{split}
\end{equation}
and
\begin{equation}
    \nabla_{\mathbf{z}^I_{i_t}} \ell = - \sum_{w \in \mathbf{w}_{i_t}} \mathbf{z}^W_w \frac{e^{-Q_{w,i_t}}}{1 + e^{-Q_{w,i_t}}} + \sum_{\tilde{w} \in \text{Neg}(\mathcal{W})}  \mathbf{z}^W_{\tilde{w}} \frac{e^{-Q_{\tilde{w},i_t}}}{1 + e^{-Q_{\tilde{w},i_t}}}.
\end{equation}

For $\mathbf{z}^I_{i_{t-j}}$, $j=1,\dots,k$, we compute their gradients as:
\begin{equation}
\label{eqn:appene}
    \begin{split}
        \nabla_{\mathbf{z}^I_{i_{t-j}}} \ell = -\mathbf{z}^I_{i_t} \frac{e^{-Q_{i,u,i_{t-1:t-k}}}}{k\big(1 + e^{-Q_{i,u,i_{t-1:t-k}}}\big)}  + \sum_{\tilde{i} \in \text{Neg}(\mathcal{I})} \mathbf{z}^I_{\tilde{i}} \frac{e^{Q_{\tilde{i},u,i_{t-1:t-k}}}}{k\big(1 + e^{Q_{\tilde{i},u,i_{t-1:t-k}}}\big)} \\
        - \sum_{w \in \mathbf{w}_{i_{t-j}}} \mathbf{z}^W_w \frac{e^{-Q_{w,i_t}}}{1 + e^{-Q_{w,i_{t-j}}}} + \sum_{\tilde{w} \in \text{Neg}(\mathcal{W})}  \mathbf{z}^W_{\tilde{w}} \frac{e^{-Q_{\tilde{w},i_{t-j}}}}{1 + e^{-Q_{\tilde{w},i_{t-j}}}}.
    \end{split}
\end{equation}
As for the embeddings for item contextual feature (word) token $w$ and users feature $x$ that show up in the observation, their gradients can be computed by:
\begin{equation}
\label{eqn:appen4}
    \nabla_{\mathbf{z}^W_w} \ell = \sum_{\substack{i: w \in i, \\ i \in \{i_t,\dots,i_{t-k}\}}} -\mathbf{z}^I_i \frac{e^{-Q_{w,i}}}{ 1 + e^{-Q_{w,i}}}
\end{equation}
and
\begin{equation}
\label{eqn:appen5}
    \nabla_{\mathbf{z}^X_x} \ell =  -\mathbf{z}^U_u \frac{e^{-Q_{x,u}}}{ 1 + e^{-Q_{x,u}}}.
\end{equation}

For the $\tilde{i}$, $\tilde{w}$ and $\tilde{w}$ that are being sampled in (\ref{eqn:appen1}) to (\ref{eqn:appen5}), their gradients can be calculated in similar fashion. We do not present them here to avoid unnecessary repetition.

\begin{algorithm}[hbt]
\caption{Framework for learning embeddings}
\label{algo:full}
\begin{flushleft}
\textbf{Input:} Training dataset $\mathcal{D}$ with $n$ observations, training epochs $e=30$; \\
\textbf{Output:} The embeddings $Z^I$, $\tilde{Z}^I$, $Z^U$, $Z^W$ and $Z^X$;
\end{flushleft}
\begin{algorithmic}[1]
\STATE Shuffle $\mathcal{D}$;
\FOR {$\text{epoch}=1,\dots,e$}
    \FOR {$q=1,\dots, n$}
        \STATE Construct negative sample sets $\text{Neg}(\mathcal{I})$, $\text{Neg}(\mathcal{W})$ and $\text{Neg}(\mathcal{X})$ by (\ref{eqn:appen-neg});
        \STATE Update item/user/word/user feature embedding of the $q^{\text{th}}$ observation according to (\ref{eqn:appen1})-(\ref{eqn:appen5}) with learning rate $\alpha_{\text{epoch}\times n + q}$ in (\ref{eqn:appen-rate});
        \STATE Update embeddings of the instances in the above negative sample sets with same learning rate.
    \ENDFOR
\ENDFOR
\end{algorithmic}
\end{algorithm}

During training, we linearly decay the learning rate after each batch update from the starting value $\alpha_0$ to 0. This practice is in accordance with other widely-used shallow embedding models such as \textsl{word2vec}, \textsl{doc2vec}, \textsl{fastText}, etc. In our synchronous stochastic gradient descent implementation, each observation constitutes its own batch. Therefore with $e$ epochs and $n$ observations, the adaptive learning rate at step $l$ is given by
\begin{equation}
\label{eqn:appen-rate}
    \alpha_l = \frac{l}{n\times e} \alpha_0 .
\end{equation}
When implementing frequency-based negative sampling, the items are sampled according to the probability computed by
\begin{equation}
\label{eqn:appen-neg}
    p(i) \propto 1 - \sqrt{\frac{10^{-5}}{f(i)}},
\end{equation}
where $f(i)$ is the frequency of item $i$ in training dataset \cite{mikolov2013distributed}. Negative sampling for words and user features are conducted in the same way. We set the number of negative samples to five under all cases in training. The pseudo-code of our algorithm is given in Algorithm (\ref{algo:full}).

\end{document}